\documentclass[a4paper,12pt,english]{article}

\usepackage{amsfonts,bm,amssymb,euscript,array,babel,cite}
\usepackage{amsmath,amsthm} 
\usepackage[dvips]{epsfig}



\newcommand{\be}{\begin{equation}} \newcommand{\ee}{\end{equation}}
\newcommand{\beq}{\begin{equation}} \newcommand{\eeq}{\end{equation}}
\newcommand{\beqa}{\begin{eqnarray}}
\newcommand{\eeqa}{\end{eqnarray}} \newcommand{\eq}[1]{(\ref{#1})}
\def\nn{\nonumber} \def\bea{\begin{eqnarray}} \def\eea{\end{eqnarray}}
\def\obar{\overline}
\def\a{\alpha}  \def\b{\beta}
  
 \def\d{\delta}

      \def\cN{{\cal N}}

\def\R{{\mathbb R}} \def\C{{\mathbb C}} \def\N{{\mathbb N}}
 \def\one{\mbox{1 \kern-.59em {\rm l}}}

\def\msu{\mathfrak{su}}

\def\bit{\begin{itemize}} \def\eit{\end{itemize}} 

\def\({\left(} \def\){\right)} \def\tens{\otimes}

\def\IR{{\hbox{{\rm I}\kern-.2em\hbox{\rm R}}}} \def\IB{{\hbox{{\rm
I}\kern-.2em\hbox{\rm B}}}} \def\IN{{\hbox{{\rm I}\kern-.2em\hbox{\rm
N}}}} \def\IC{\,\,{\hbox{{\rm I}\kern-.59em\hbox{\bf C}}}}
\def\IZ{{\hbox{{\rm Z}\kern-.4em\hbox{\rm Z}}}} \def\IP{{\hbox{{\rm
I}\kern-.2em\hbox{\rm P}}}} \def\IH{{\hbox{{\rm I}\kern-.4em\hbox{\rm
H}}}} \def\ID{{\hbox{{\rm I}\kern-.2em\hbox{\rm D}}}}
\def\II{{\hbox{\rm I}\kern-.2em\hbox{\rm I}}}

 \allowdisplaybreaks[3]

\begin{document}

\renewcommand{\title}[1]{\vspace{10mm}\noindent{\Large{\bf
#1}}\vspace{8mm}} \newcommand{\authors}[1]{\noindent{\large
#1}\vspace{5mm}} \newcommand{\address}[1]{{\itshape #1\vspace{2mm}}}

\begin{titlepage}
\begin{flushright}
UWThPh-2007-15
\end{flushright}

\begin{center}

\title{ \Large Fermions on \\[1ex] spontaneously generated 
spherical extra dimensions}

\vskip 3mm

\authors{Harold {\sc Steinacker}}

\vskip 3mm

\address{Faculty of Physics, University of Vienna\\
Boltzmanngasse 5, A-1090 Wien, Austria\\ E-mail:
harold.steinacker@univie.ac.at}

\vskip 3mm

\authors{George {\sc Zoupanos}}

\address{ Physics Department National
Technical University\\ Zografou Campus, GR-15780 Athens\\
E-mail:George.Zoupanos@cern.ch}

\vskip 1.4cm

\textbf{Abstract}

\vskip 3mm

\begin{minipage}{14cm}%

We include fermions to the model proposed in 
 {\tt hep-th/0606021}, and obtain a renormalizable 4-dimensional $SU(N)$ 
gauge theory which spontaneously generates fuzzy extra dimensions
and behaves like Yang-Mills theory on $M^4 \times S^2$.
We find a truncated tower of fermionic Kaluza-Klein states
transforming under the low-energy gauge group, which 
is found to be either $SU(n)$, or $SU(n_1) \times SU(n_2) \times U(1)$.
The latter case implies a nontrivial $U(1)$ flux on $S^2$,
leading to would-be zero modes for the bifundamental fermions. 
In the non-chiral case they may 
pair up to acquire a mass, and the emerging picture is that
of  mirror fermions. 
We discuss the possible implementation of a chirality constraint
in 6 dimensions, which is nontrivial at the quantum level due to the fuzzy nature 
of the extra dimensions.

\end{minipage}

\end{center}

\end{titlepage}

 \tableofcontents

\section{Introduction}

The idea of unification of interactions in higher dimensions 
is central for many modern developments in the theory of elementary
particles and fields, going back to Kaluza-Klein. 
Recently, a surprising new twist has entered this programme: 
It was found that extra dimensions can arise effectively
within a 
4-dimensional renormalizable gauge theory, as an effective description valid up to some energy scale. 
This has become known under the name of deconstruction 
\cite{Arkani-Hamed:2001ca}. 

A strikingly simple realization of the idea of a spontaneous
generation of  extra dimensions was given in
\cite{Aschieri:2006uw}, inspired 
by an earlier work \cite{Aschieri:2003vy}.
Since we will extend this model here, we
briefly recall the main features of \cite{Aschieri:2006uw}.
The model is simply $G=SU(\cN)$ Yang-Mills theory on $M^4$
for some generic
(large) $\cN \in \N$, with 3 scalars in the adjoint of 
$G$ transforming
as vectors under a global $SO(3)$ symmetry. It turns out that 
adding the most general renormalizable potential leads 
to SSB and to
the formation of an extra-dimensional fuzzy sphere
via the Higgs effect. The unbroken gauge group is generically
$K = SU(n_1) \times SU(n_2)
\times U(1)$, or possibly $K = SU(n)$. 
The gauge fields on $S^2_N$ arise 
from fluctuations of the $\msu(\cN)$-valued
scalar fields which form the extra-dimensional sphere.
For energies less than $\Lambda_{6D} =\frac{N^2}R$,
the appropriate description of the
model is then as Yang-Mills theory with gauge group $K$ on $M^4
\times S^2_N$. Here $R$ is the
radius of the internal fuzzy sphere, which is determined 
(along with the other low-energy
 parameters including $n_1,n_2$) by the
coupling constants of the model. 
This interpretation was confirmed
by the full harmonic analysis, i.e. by recovering precisely the
expected Kaluza-Klein modes, up to the cutoff $\Lambda_{6D}$. Above
that energy scale, the model again behaves like a 4D  gauge
theory, thus maintaining  renormalizability. 
The main features of compactification on higher dimensions are
hence realized
within the framework of renormalizable 4D field theory.

This dynamical or spontaneous generation of extra dimensions is of
course strongly suggestive of gravity. Indeed, the results of
\cite{Steinacker:2007dq} allow to understand this mechanism 
in terms of gravity: the scalar potential defines a matrix-model
action which - using a slight generalization of \cite{Steinacker:2007dq} -
can be interpreted as nonabelian Yang-Mills coupled to dynamical 
Euclidean gravity in the extra dimensions.

In the present paper, we add fermions to this model, and 
work out their effective description from both the 6D and 4D 
point of view. In particular, we
show how to obtain a model which has an effective description as
Yang-Mills theory on $M^4 \times S^2_N$, with fermions
coupling appropriately to the 6D gauge fields
and transforming under the unbroken gauge group
$SU(n_1) \times SU(n_2)
\times U(1)$ resp. $SU(n)$. 

In order to make renormalizability manifest, 
we start again from the 4D point of view, and add fermions transforming
appropriately under the symmetries of the bosonic sector. 
Renormalizability strongly restricts 
the possible Yukawa couplings between
the fermions and the scalar fields.
We then determine whether the fermions acquire the expected 
action for the effective 
6-dimensional space $M^4 \times S^2_N$. 

We first show in section \ref{sec:Weylfermions} 
that adding a ``minimal'' set of fermions does not
lead to the desired 6D behavior. 
However upon doubling the set of fermions, the appropriate 
6D picture is indeed found. 
As shown in detail in section \ref{sec:double-fermions}, 
the effective description is that of 
Dirac fermions on $M^4 \times S^2_N$. 
This is confirmed by explicitly 
identifying all Kaluza-Klein modes on $S^2_N$, and determining their
masses in the effective 4-dimensional description.

The (generic) case of the low-energy gauge group 
$SU(n_1) \times SU(n_2)\times U(1)$ is particularly interesting. 
The extra-dimensional sphere then automatically  
carries a magnetic flux, which couples to 
the fermions transforming
in the bifundamental of $SU(n_1) \times SU(n_2)$. 
According to the index theorem,
this implies that these fermions have zero modes, which
are expected to become precisely the 
massless fermions from the 
4D point of view. This conclusion is only true for a 
chiral 6D theory; in the non-chiral case, 
two such ``would-be zero modes'' with opposite chirality can form a 
massive Dirac fermion. 

We study the above mechanism in the present model in section 
\ref{sec:type2-zeromodes}. The expected 
(would-be) zero modes are indeed found, in agreement with the 
theoretical expectations. However since the fermions behave like 
Dirac fermions on $M^4 \times S^2_N$, these would-be zero modes
indeed acquire a mass unless some fine-tuning is imposed.
One would therefore like to impose a chirality constraint
on the fermions. 
This is difficult here, because the chirality
operator on the fuzzy sphere is a dynamical operator 
depending on the scalar fields. While a chirality constraint
can be imposed on the classical level, its implementation on the
quantum level is not clear. Therefore we arrive at a
picture of ``mirror fermions'', where each chiral fermion
has a partner with opposite chirality and quantum numbers.
Such models have been considered from a phenomenological 
point of view in \cite{Maalampi:1988va}.

The model shows some intriguing features hinting at 
a simpler structure at high energies. In particular, we discuss in 
section \ref{sec:SU2N} an extended $SU(2\cN)$ structure which 
naturally accommodates both the bosonic and the fermionic matter.
It also suggests a natural way of obtaining a chiral model, 
using essentially projector-valued fields. Nevertheless 
its consistency at the quantum level 
(i.e. renormalizability) is not clear, and at present 
it is meant mainly as a stimulation for further research. 

There are many possible generalizations and variants of the 
model discussed here. In particular, we discuss in section 
\ref{sec:fluxons-ssb} a possible mechanism for 
further symmetry breaking
using so-called fluxons on $S^2_N$, which are 
non-classical, topologically nontrivial
solutions of gauge theory on $S^2_N$. Generalizations to other 
fuzzy internal spaces may allow to obtain chiral models. 
It is also interesting here to recall the analysis of 
\cite{Dolan:2002ck}, 
where the spectrum of the standard model has been
related to the zero modes of the Dirac operator on other fuzzy spaces;
the generation of a nontrivial index on $S^2_N$
has also been discussed in \cite{Aoki:2004sd,Aoki:2006wv}.
Finally, a similar supersymmetric model for spherical deconstruction
has already been given in \cite{Andrews:2005cv}. However, the
remarkable mechanism in our model 
for selecting a single vacuum with particular
unbroken gauge group out of the vast number of possibilities 
 is lost there, and a SUSY version preserving this mechanism would 
be very desirable.

\section{The bosonic action}

We start by recalling the definition and main features of the model
in \cite{Aschieri:2006uw}. 
Consider the $SU(\cN)$ gauge theory on 4-dimensional Minkowski
space $M^4$ with coordinates $y^\mu$, $\mu = 0,1,2,3$, with action
\be 
{\cal S}_{YM}= \int d^{4}y\, Tr\,\left(
\frac{1}{4g^{2}}\, F_{\mu \nu}^\dagger F_{\mu \nu} +
(D_{\mu}\phi_{{a}})^\dagger D_{\mu}\phi_{{a}}\right) - V(\phi) .
\label{the4daction}
\ee 
Here $A_\mu$ are $\msu(\cN)$-valued gauge fields, $D_\mu =
\partial_\mu + [A_\mu,.]$, and 
\be \phi_{{a}} = - \phi_{{a}}^\dagger
~, \qquad a=1,2,3 
\ee 
are 3  traceless antihermitian scalars in the
adjoint of $SU(\cN)$,
\be 
\phi_{{a}} \to U^\dagger \phi_{{a}} U ,
\ee
where $U = U(y) \in SU(\cN)$. Furthermore, the $\phi_a$ transform as
vectors of an additional global $SO(3)$ resp. $SU(2)$ symmetry. 
$V(\phi)$ is of course the most general renormalizable potential
invariant under the above symmetries, which can be written as
\begin{eqnarray}
V(\phi) &=& Tr\, \left( g_1 \phi_a\phi_a \phi_b\phi_b +
g_2\phi_a\phi_b\phi_a \phi_b - g_3 \varepsilon_{a b c} \phi_a \phi_b
\phi_c + g_4\phi_a \phi_a \right) \nn\\ && + \frac{g_5}{\cN}\,
Tr(\phi_a \phi_a)Tr(\phi_b \phi_b) + \frac{g_6}{\cN} Tr(\phi_a
\phi_b)Tr(\phi_a \phi_b) \label{pot} \\[1ex]
&=& Tr \( a^2(\phi_a\phi_a + \tilde b\, \one)^2  +\frac 1{\tilde g^2}\,
F_{ab}^\dagger F_{ab}\,\) + \frac{h}{\cN}\, g_{ab} g_{ab}
\label{V-general-2}
\end{eqnarray}
for suitable constants $a,b,\tilde g,h,$ dropping a constant
shift. Here 
\bea 
F_{{a}{b}}
&=& [\phi_{{a}}, \phi_{{b}}] - \varepsilon_{abc} \phi_{{c}}\, =
\varepsilon_{abc} F_c , \nn\\ \tilde b &=& b + \frac{d}{\cN} \,
Tr(\phi_a \phi_a), \qquad 
g_{ab} = Tr(\phi_a \phi_b).
\label{const-def}
\eea 
We also performed a rescaling 
\be \phi'_a = R\; \phi_a, \qquad R = \frac{2 g_2}{g_3} ,
\label{Radius}
\ee 
where $R$ has dimension of length; we will usually suppress $R$ and
drop the prime.  
Here $\tilde b =\tilde b(y)$ is a scalar field, 
$g_{ab} = g_{ab}(y)$ is a symmetric
tensor field under the global $SO(3)$, and $F_{ab}=F_{ab}(y)$ is an
$\msu(\cN)$-valued antisymmetric tensor field which will be
interpreted as field strength on the spontaneously generated fuzzy
sphere.  
In this form, $V(\phi)$ looks indeed like the action of
Yang-Mills gauge theory on a fuzzy sphere $S^2_N$
\cite{Steinacker:2003sd,Carow-Watamura:1998jn,
Presnajder:2003ak}. In particular, the term
$(\phi_a\phi_a + \tilde b)^2$ is
 necessary for the interpretation as 
a pure YM action on $S^2_N$
 involving only tangential gauge fields, and it
determines and stabilizes a unique vacuum.

It is easy to see that at one loop, 
the parameters $R, a$, $\tilde g$, $d$ and $h$
are logarithmically divergent, 
while $b$ and therefore $\tilde b$ is quadratically divergent.
The gauge coupling $g$ is asymptotically free. 
A full analysis of the RG flow of these parameters is complicated by
the fact that the vacuum and the number of massive resp. massless
degrees of freedom depends sensitively on the values of these
parameters, with
different effective description at different energy scales.
This will be discussed next.

\subsection{The minimum of the potential and SSB}
\label{sec:vacua}

The mechanism for the generation (or deconstruction) 
of extra dimensions 
in this model is based on spontaneous symmetry breaking 
and the ordinary Higgs effect. 
We first have to determine the vacuum, i.e. the minimum of $V(\phi)$. 
This vacuum turns out to have a geometric interpretation as
$M^4 \times S^2_N$, breaking $SU(\cN)$ down to a smaller gauge group.
The geometric interpretation is confirmed using harmonic analysis, 
i.e. identification of the Kaluza-Klein (KK) modes.
The Higgs effect then induces the appropriate masses of the
higher Kaluza-Klein modes of the fuzzy sphere $S^2_N$.

To determine the minimum of $V(\phi)$
\eq{V-general-2} turns out to be a rather nontrivial task, and
the answer depends crucially on the parameters in the potential.
The potential is positive
definite provided 
\be a^2 >0, \qquad \frac 2{\tilde g^2} \, >0, \qquad h \geq 0, 
\ee
which we assume in the following.  
For suitable values of the parameters in the potential, we can
immediately write down the vacuum. 
Assume  $h=0$  for simplicity. Since
$V(\phi) \geq 0$, the global minimum of the potential is then certainly
achieved if 
\be F_{{a}{b}} = [\phi_{{a}}, \phi_{{b}}] -
\varepsilon_{abc} \phi_{{c}}~ = 0, \qquad -\phi_a\phi_a = \tilde b,
\label{vacuum-trivial-cond}
\ee 
because then $V(\phi) =0$. This implies that $\phi_a$ is a
representation of $SU(2)$, with prescribed Casimir\footnote{note that
$-\phi\cdot \phi = \phi^\dagger\cdot \phi >0$ since the fields are
antihermitian} $\tilde b$. These equations may or may not have a
solution, depending on the value of $\tilde b$.  Assume first that
$\tilde b$ coincides with the quadratic Casimir of a
finite-dimensional irrep of $SU(2)$, 
\be 
\tilde b = C_2(N) = \frac 14(N^2-1) 
\ee 
for some $N \in \N$. If furthermore the dimension $\cN$ of
the matrices $\phi_a$ can be written as \be \cN = N n, \ee then
clearly the solution of \eq{vacuum-trivial-cond} is given by 
\be
\phi_a = X_a^{(N)} \otimes \one_{n}
\label{vacuum-trivial}
\ee 
up to a gauge transformation, where $X_a^{(N)}$ denote the
generator of the $N$-dimensional irrep of $SU(2)$. This can be
viewed as a special case of \eq{solution-general} below, consisting of
$n$ copies of the irrep $(N)$ of $SU(2)$.

For generic $\tilde b$, \eq{vacuum-trivial-cond} cannot
be satisfied. The exact
vacuum (which certainly exists since the potential is positive
definite) can  be found by solving the ``vacuum equation''
$\frac{\d V}{\d \phi_a} =0$, 
\be 
a^2 \{\phi_a,\phi\cdot\phi + \tilde b
+ \frac{d}{\cN}\, Tr(\phi\cdot \phi+\tilde b)\} + \frac{2h}{\cN}
g_{ab}\phi_b + \frac 1{\tilde g^2}\, (2[F_{ab},\phi_b] + F_{bc}
\varepsilon_{abc}) =0 ,
\label{eom-int}
\ee 
where $\phi\cdot\phi \equiv \phi_a \phi_a$.

The general solution of \eq{eom-int} is not known. 
However, it is easy
to write down a large class of solutions: any decomposition of $\cN =
n_1 N_1 + ... + n_h N_h$ into irreps of $SU(2)$ with multiplicities
$n_i$ leads to a block-diagonal solution 
\be \phi_a = diag\Big(\a_1\,
X_a^{(N_1)}, ..., \a_k\, X_a^{(N_k)}\Big)
\label{solution-general}
\ee 
of the vacuum equations \eq{eom-int}, where $\a_i$ are suitable
constants which are determined by the equations of motion. 
We can expect that this Ansatz indeed contains the true vacuum
at least  for a reasonable range of parameters,
because it is known to reproduce all standard ``commutative'' 
solutions of YM on $S^2$ \cite{Steinacker:2003sd,Steinacker:2007iq}. It turns out that 
 only 2 cases occur:

\paragraph{Type I vacuum.}

Let $N$ be the dimension of the irrep whose
Casimir $C_2(N)\approx \tilde b$ is closest to $\tilde b$. If
furthermore the dimensions match as $\cN = N n$, we expect that the
vacuum is given by $n$ copies of the irrep $(N)$, which can be written
as 
\be \phi_a = \a\, X_a^{(N)} \otimes\one_{n}.
\label{vacuum-mod1}
\ee 
This is a slight generalization of \eq{vacuum-trivial}, with $\a$
being determined through the vacuum equations \eq{eom-int}.
A vacuum of the form \eq{vacuum-mod1} will be denoted as ``type I
vacuum''.  As explained in detail in \cite{Aschieri:2006uw}, it should
be interpreted as a spontaneously generated extra-dimensional
fuzzy sphere $S^2_{N}$, where $x_a \sim \frac 1N\, X_a^{(N)}$ are the
coordinates of
$S^2_{N}$ \eq{fuzzycoords}. This is confirmed using 
harmonic analysis, i.e. by decomposing all fields into the 
correct harmonics on $M^4 \times S^2_N$.

\paragraph{Type II vacuum.} 

In the generic case, the vacuum is expected to
consist of several distinct blocks. This  necessarily happens 
if $\cN$ is not divisible by the dimension of the irrep whose
Casimir is closest to $\tilde b$. Assuming that 
$\tilde N = \sqrt{4\tilde b +1}$ is large, 
it was shown in  \cite{Aschieri:2006uw} that
the solution with minimal potential among all possible
partitions \eq{solution-general} is given either by a type I vacuum, or
takes the form 
\be \phi_a
= \left(\begin{array}{cc}\a_1\, X_a^{(N_1)}\otimes\one_{n_1} & 0 \\ 0
& \a_2\,X_a^{(N_2)}\otimes\one_{n_2}
             \end{array}\right),
\label{vacuum-mod2}
\ee 
as long as the integers $N_1,N_2$ satisfy 
$\frac{N_i}{\tilde N} \approx 1$ and of course 
$\cN = N_1 n_1 + N_2 n_2$. Furthermore, the vacuum turns out to satisfy
\be 
N_2 = N_1+1 ,
\ee
with uniquely determined $N_i$ and $n_i$. 
A vacuum of the form \eq{vacuum-mod2} will be denoted as ``type II
vacuum'', which is the generic case.
Using a rather robust convexity argument \cite{Aschieri:2006uw}, 
one can show 
that more than 2 different types of blocks $N_i$ do not occur for the
vacuum.

\paragraph{6D interpretation}

As shown in  \cite{Aschieri:2006uw}, the fluctuations of the 
scalars or ``covariant coordinates''
\be
\phi_a = \a X_a + A_a
\ee
 together withe the gauge fields
$A_\mu$ provide the components of a 6D gauge field 
$A_M = (A_\mu, A_a)$ on $M^4 \times S^2_N$.  The effective action from a 6D point of
view is  that of Yang-Mills on  $M^4 \times S^2_N$ with gauge group
$SU(n)$
for the type I vacuum, and with gauge group $SU(n_1) \times SU(n_2)
\times U(1)$ for the type II vacuum.
The latter comes with an induced $U(1)$ magnetic monopole on $S^2_N$
with monopole number $k = N_1 - N_2$, hence $k=1$ according to the
above analysis. The radial components of the fields $\phi_i$ on $S^2$
are very massive due to the term $(\phi_a\phi_a + \tilde b\, \one)^2$,
and not visible at low energies. 

\paragraph{Fluxons}

There is a further type of solutions to the equations of motion \eq{eom-int},
known as fluxon in the context of noncommutative gauge theory. It is
given by (one or several) one-dimensional blocks of the form
\be
\phi_a = c_a \quad \in\, i\, \R
\ee
corresponding to a vector $\vec c \in \R^3$. Its length $\vec c^2 = \sum_a c_a^2
\approx -\tilde b$ is determined by the e.o.m., minimizing the
potential. Since the term $(\phi_a\phi_a + \tilde b\, \one)^2$ dominates
assuming that $a^2 \approx \frac 1{\tilde g^2}$, 
such a fluxon block contributes typically
$S_{fluxon} \approx \frac 1{\tilde g^2}\, \vec c^2 
\approx \frac 1{\tilde g^2}\,\tilde N^2 $ to the action through the
field strength. 
This is large compared to the ``regular'' solutions of type I and type
II vacuum, and was therefore not considered 
any further in \cite{Aschieri:2006uw}.
Nevertheless this may play a role for relatively small $\cN$,
and the possibility of 
off-diagonal terms as discussed in section 
\ref{sec:fluxons-ssb} justifies 
further consideration. There exist further solutions to 
 the equations of motion \eq{eom-int}, which are however
strongly suppressed and not expected to be relevant here.

\section{Fermions}
\label{sec:fermions}

We now want to add fermions to our model \eq{the4daction}. 
In order to ensure renormalizability we
start with the 4-dimensional point of view, 
and  write down the most general renormalizable Lagrangian
compatible with the symmetries. The fermion content and their
transformation under the symmetries of the above model
are chosen such that they have a chance to behave like 
6-dimensional fermions in the vacuum corresponding to
$M^4 \times S^2$. 

Let us briefly summarize the main steps.
We start with the minimal
case of adding 
4D Weyl spinors in the adjoint of $SU(\cN)$
which transform as a doublet of the global $SU(2)$ symmetry.  
However, it turns out that no 6-dimensional behavior is found,
more precisely no kinetic term arises in the extra dimensions.

This problem will be cured by adding 
a second doublet of 4D Weyl spinors. The most general 
renormalizable Yukawa interaction then naturally leads to the 
Dirac operator on the spontaneously  generated fuzzy sphere, 
and the effective description is 
indeed that of a Dirac fermion on 
$M^4 \times S^2_N$. In fact, the requirement of 
renormalizability uniquely singles out the ``standard'' 
Dirac operator on $S^2_N$ \cite{Grosse:1994ed} rather than any of 
the other candidates that 
have been proposed in the literature.

The KK-modes and 
the low-energy properties of the fermions depend of course
on the vacuum. In a type I vacuum, the fermions live in the 
adjoint of the unbroken $SU(n)$ gauge group,
and no zero modes are found.
In a type II vacuum, the fermions 
couple to the unbroken 
$SU(n_1) \times SU(n_2) \times U(1)$ gauge group, and
the off-diagonal block components 
$\Psi^{12}$ and $\Psi^{21}$ then transform in the bifundamental 
$(n_1) \times (\obar n_2)$ resp. $(\obar n_1) \times ( n_2)$
of $SU(n_i)$ and $SU(n_2)$, with opposite charge under
the $U(1)$. Therefore they feel the $U(1)$
magnetic monopole which is induced in that vacuum
\cite{Steinacker:2003sd}, with opposite charge. 
The index theorem then applies, and guarantees
the existence of ``would-be zero modes'' 
for the chiral components
of $\Psi^{12}$ and $\Psi^{21}$. 
Nevertheless, they may pair up and acquire a mass 
because the model is non-chiral. 

Imposing a chirality constraint corresponding to
chiral fermions on $M^4 \times S^2_N$ turns 
out to be difficult. 
On the level of the effective action, we discuss 2 possible 
chirality constraints, which imply 
the existence
 of $k$ exact chiral zero modes\footnote{the dynamically preferred vacuum
was shown to have $k=1$ in \cite{Aschieri:2006uw}} as 
expected in a background 
with magnetic charge $k$. 
However, the problem is that the chirality operators on $S^2_N$
necessarily contain the dynamical fields $\phi_a$ which define 
the extra dimensions, and this operator has a clear meaning only 
in or near the geometric vacua. Therefore the implementation
of such a chirality constraint on the quantum level 
is highly nontrivial,
and we are not able to define a renormalizable model 
which describes {\em chiral} fermions on $M^4 \times S^2_N$.
Accordingly we have a doubling of modes, and
the would-be zero modes may pair up to become 
massive Dirac fermions from the low-energy point of view.
This leads to a situation analogous to the ``mirror fermions''
 \cite{Maalampi:1988va}.

\paragraph{The commutative case: fermions on $M^4 \times S^2$}

We first recall the classical description of 
fermions on $M^4 \times S^2$, formulated in a way which will
generalize to the fuzzy case. This is done using the embedding
$S^2 \hookrightarrow \R^3$ based on the 7-dimensional Clifford algebra
\be
\Gamma^A = (\Gamma^{\mu},\Gamma^a) = (\one \otimes \gamma^\mu,
\sigma^a \otimes i\gamma_5) .
\label{gamma-6D-0}
\ee
Here $\sigma^a,\,\,a=1,2,3$ generate the 2-resp. 3-dimensional
Clifford algebra.
The $\Gamma^A$ act on $\C^2 \otimes \C^4$  and satisfy
$(\Gamma^A)^\dagger = \eta^{AB} \Gamma^B$ where 
$\eta^{AB} = (1,-1,...,-1)$ is the 7-dimensional Minkowski metric.
The corresponding 8-component spinors describe
Dirac fermions on $M^4 \times S^2$, and can be viewed as 
Dirac spinors on $M^4$ tensored with 
2-dimensional Dirac spinors on $S^2 \hookrightarrow \R^3$. 
We can define a 2-dimensional chirality operator 
$\chi$ locally at each point of the unit sphere $S^2$ by setting 
\be
\chi = x_a \sigma^a ,
\ee
which has eigenvalues $\pm 1$. At the north pole
$x_a = (0,0,1)$ of $S^2$ this coincides with $\chi = -i \sigma^1
\sigma^2 = \sigma^3$,
as expected. This can be understood as usual 
in terms of a comoving frame, adding a unit vector which is
perpendicular to $S^2$.
The action for a Dirac fermion on $M^4 \times S^2$  can then be written
as
\be 
S_{6D} = \int\limits_{M^4} d^4 y\, \int\limits_{S^2} d\Omega\,\,\obar\Psi_D \left(
i \gamma^\mu \partial_\mu 
+ i\gamma_5  \not \!\!D_{(2)} + m \right)\Psi_D ,
\ee
where
\be
\not \!\! D_{(2)}\Psi_D = (\sigma_a L_a  + 1) \Psi_D 
\label{2Ddirac-class}
\ee
is the Dirac operator on $S^2$ in ``global'' notation. Here
$L_a = i \varepsilon_{abc} x_b \partial_c$ is the angular momentum
operator,
and the constant 1 in \eq{2Ddirac-class}  ensures 
$\{\not \!\!\! D_{(2)},\chi\}=0$ and reflects the curvature of $S^2$.
This is 
equivalent to the standard formulation in terms of a comoving frame,
but more appropriate for the fuzzy case. 

Chiral (Weyl) spinors $\Psi_\pm$ on  $M^4 \times S^2$ are then defined  using the
6D chirality operator
\be
\Gamma =  \gamma_5 \chi ,
\label{chiral-6D-class}
\ee
and satisfy $\Gamma \Psi_\pm = \pm \Psi_\pm$.
They contain both chiralities from the 4D point of view, 
\be
\Psi_\pm = (0,1;\pm) + (1,0;\mp) ,
\ee
where $(0,1;\pm)$ denotes a Weyl spinor $\psi_\a$ on $M^4$ with
eigenvalue $\pm 1$ of $\chi$, and $(0,1;\mp)$  a dotted Weyl spinor 
$\obar\psi^{\dot\a}$ on $M^4$ with eigenvalue $\mp 1$ of $\chi$.
These components are of course mixed under the 6-dimensional rotations.

Majorana spinors on $M^4 \times S^2$ satisfy
$\Psi^* = C \Psi$  
where $C$ is the 6D charge conjugation operator given by
\be
C = i \gamma_2 \sigma_2
\ee
which satisfies
\be
C \Gamma^A C^{-1} = -(\Gamma^A)^* .
\ee

\subsection{Minimal Weyl fermions.}
\label{sec:Weylfermions}

Consider now our 4-dimensional model \eq{the4daction}, and let us
try to include a doublet of chiral 4-dimensional Weyl spinors
\be
\Psi(y) = \left(\begin{array}{c} \psi_{1,\a}(y) \\ 
               \psi_{2,\a}(y) \end{array}\right),
\ee
which transforms in the fundamental representation of the global
$SU(2)$ acting on the index $i$.
Since we want them to behave like spinors on 
$M^4 \times S^2_{N}$, the $\psi_{i,\a}(y)$ must be
$\cN \times \cN$ matrices; furthermore,
since the kinetic term on 
$S^2_N$ can arise in our model
only through commutators $[\phi_i,.]$ 
they must transform in the adjoint of $SU(\cN)$. 
In particular, the anomaly then vanishes. 
Fermions in the fundamental of $SU(\cN)$ are therefore not
considered here.

Thus the $\psi_{1,2}(y)$ are (Grassmann-valued) Weyl
spinors  which transform under
$SU(\cN)$ as $\psi_i(y) \to U(y)^\dagger \psi_i(y) U(y)$.
Then the kinetic term of the action is 
\be
S_K = \int d^4 y\, Tr\, \Psi^\dagger 
i \gamma^\mu (\partial_\mu + [A_\mu,.]) \Psi
 =  \int d^4 y\, Tr\, (\psi_{i,\a})^\dagger 
i ({\obar\sigma}^\mu)^{\dot\a\b} (\partial_\mu + [A_\mu,.]) \psi_{i,\b} 
\label{kinetic-fermions}
\ee
which is invariant under all the symmetries.
 It is easy to check
that the gauge sector is asymptotically free.
Furthermore we should add mass terms and Yukawa couplings,
which will lead to Dirac and chirality 
operators on the fuzzy internal 
space $S^2_{N}$. Renormalizability excludes terms 
with more than one scalar field $\phi_a$. 
To preserve Lorentz invariance, these term must include 2 
unconjugated (or 2 conjugated) spinors.  The 
only possible mass term is
\be
S_m = \int d^4 y\,  Tr\, \psi_{i,\a} 
\varepsilon^{\a\b} \varepsilon^{i j} m \psi_{j,\b}
 + h.c. \quad \equiv 0 ,
\label{majorana-mass}
\ee
which vanishes due to the Grassmann nature of the
spinors (this will no longer true once we double the fermions in
Section \ref{sec:double-fermions}).
However, there exist a
non-trivial renormalizable Yukawa interaction
\be
S_Y  = \int d^4 y\,  Tr\, \psi_{i,\a} 
\varepsilon^{\a\b}\varepsilon^{i j} 
{(\sigma_a)}^{jk} \phi_a\,  \psi_{k,\b} \, + h.c.
\label{S-yukawa-1}
\ee
Using 
\be
\varepsilon^{ij} {(\sigma_a)_j}^{k} =  \varepsilon^{kj} {(\sigma_a)_j}^{i}
\ee
it follows that
\be
Tr\, \psi_{i,\a} 
\varepsilon^{\a\b}\varepsilon^{i j} {(\sigma_a)_j}^{k} 
\phi_a \psi_{k,\b}' 
=  Tr\,  \psi_{k,\b}' \varepsilon^{\b\a}\varepsilon^{kj} 
{(\sigma_a)_j}^{i} \psi_{i,\a} \phi_a .
\label{S-yuk-id}
\ee
Therefore \eq{S-yukawa-1} is in fact the most general 
renormalizable Yukawa interaction, which can be written as 
\be
S_Y  = \frac 12 \int d^4 y\,  Tr\, \psi_{i,\a} 
\varepsilon^{\a\b}\varepsilon^{i j}
{(\sigma_a)_j}^{k}  \{\phi_a, \psi_{k,\b} \}  \, + h.c. 
\label{S-yukawa}
\ee
This involve the fuzzy chirality operator \eq{chi-def-1}
on $S^2_N$
as discussed below.  On the other hand, the analogous term
involving the fuzzy Dirac operator \eq{fuzzydirac}
on $S^2_N$ vanishes,
\be
 Tr\, \psi_{i,\a} \varepsilon^{\a\b}\varepsilon^{i j} 
{(\sigma_a)_j}^{k}  [i\phi_a, \psi_{k,\b}]  \equiv 0.
\label{dirac-vanishes}
\ee
Therefore no 6-dimensional behavior is found. This will
be cured in the next section.

\subsubsection{Dirac operator and chirality on the fuzzy sphere.}

We collect here the main facts about the  
``standard'' Dirac operator on the fuzzy sphere \cite{Grosse:1994ed},
which is given by the following analog of \eq{2Ddirac-class}
\be
\not \!\! D_{(2)}\Psi = \sigma_a [i X_a,\Psi] +  \Psi ,
\label{fuzzydirac-0}
\ee
where $[X_a,X_b] = \varepsilon_{abc} X_c$ generate the fuzzy sphere
as explained in appendix 2; recall that 
$X_a$ is antihermitian here. $\not \!\! D_{(2)}$ acts on 2-component spinors
\be
\Psi = \left(\begin{array}{c} \psi_{1} \\ 
               \psi_{2} \end{array}\right).
\ee
For spinors in the adjoint of the gauge group, the generators $X_a$
are replaced by the covariant coordinates $\phi_a$, and 
the gauged Dirac operator is
\bea
\not \!\! D_{(2)}\Psi &=& \sigma_a [i\phi_a,\Psi] +  \Psi \nn\\
&=& \sigma_a [i\phi_a,\Psi] + \{i\phi_0,\Psi\}.
\label{fuzzydirac}
\eea
Here we introduce $\phi_0 \equiv -\frac i2$ for later convenience.
This operator will arise automatically in section 
\ref{sec:double-fermions}, singled out
from other possible fuzzy Dirac operators
\cite{Carow-Watamura:1996wg,Balachandran:2000du,Balachandran:2003ay} 
by the requirement 
of renormalizability. For the time being we focus on the simplest case \eq{vacuum-mod1}.

There exists no chirality operator which anticommutes 
with $\not \!\! D_{(2)}$ and has
eigenvalues $\pm 1$; this follows from the spectrum of $\not \!\!
D_{(2)}$, which will be determined below \eq{D-spectrum}.
Nevertheless, 
there is a clear notion of approximate chirality for a given vacuum  
at least for the low-lying modes: 
consider the covariant operator  \cite{Grosse:1994ed}
\bea
\chi(\Psi) &=& \frac 1{N}\,\sigma_a \{i\phi_a,\Psi \} 
\label{chi-def-1}\\
&=& \frac 1{N}\,(\sigma_a \{i\phi_a,\Psi \} + [i\phi_0,\Psi])
\eea
which is  invariant under the global $SU(2)$.
To see this, the extended notation  
\be
\Phi = \phi_a \sigma^a + \phi_0 \sigma^0
\ee
of section \ref{sec:SU2N} is convenient. Then
using
\bea
(N\chi + \not \!\! D_{(2)})\Psi &=& 2i(\sigma_a \phi_a + \phi_0) \Psi = 2i \Phi \Psi, \label{Phi-chi-D-relation-1}\\
(N\chi - \not \!\! D_{(2)})\Psi &=& 2i(\sigma_a  \Psi\phi_a -\Psi\phi_0)
\label{Phi-chi-D-relation-2}
\eea
it follows that
\be
2N(\not \!\! D_{(2)}\chi + \chi \not \!\! D_{(2)}) \Psi 
= - 4[(\phi_a\phi_a + \phi_0 \phi_0,\Psi] 
- 2 i \sigma_a \varepsilon_{abc} \{F_{bc},\Psi\}
\ee
and thus
\be
(\not \!\! D_{(2)}\chi + \chi \not \!\! D_{(2)}) \Psi
= - \frac iN\, \sigma_a \varepsilon_{abc} \{F_{bc},\Psi\}
- \frac 2N\, [(\phi_a\phi_a + \phi_0 \phi_0),\Psi] 
\label{D-chi-anticomm-gen}
\ee
which is approximately zero, and exactly zero for $F=0$.
Moreover, $\chi^2 \approx -\frac 4{N^2} \phi_a^2 \propto \one$ 
using \eq{chi-def-1}, 
at least for low modes.
Therefore $\chi$ plays the role of a chirality operator on the fuzzy
sphere \cite{Grosse:1994ed}. 
This can be understood by considering e.g. 
the north pole $(x_1 \approx x_2 \approx 0, x_3\approx R)$ of $S^2_N$, 
where the tangential Clifford algebra
is generated by $\sigma_1$ and $\sigma_2$; then 
$\chi \approx i\sigma_1 \sigma_2 =  \sigma_3$. 
In particular, \eq{Phi-chi-D-relation-1} 
implies that for low modes, $\chi$ can be replaced by
\be
\chi \Psi \approx \frac {2i}N \Phi \Psi.
\label{Psi-left-chi}
\ee
The rhs of \eq{Psi-left-chi} thus provides an
interesting  alternative definition of chirality
on $S^2_N$, which is related but not identical 
to the Ginsparg-Wilson approach
\cite{Balachandran:2003ay}. 

The relation \eq{D-chi-anticomm-gen} 
implies as usual that the eigenvalues $E_{n,\pm}$ 
come in pairs with opposite sign, except for simultaneous 
eigenvectors of 
of $\chi$ and $\not \!\! D_{(2)}$ where
either $\chi$ or $\not \!\! D_{(2)}$ vanish. Indeed, note that
$\sigma_a \varepsilon_{abc} \{F_{bc},\Psi\} \propto \chi \Psi$
for any of the vacua under consideration here; therefore
$\chi\Psi$ is an eigenvector of $\not \!\! D_{(2)}$ for
any eigenvector $\Psi$ of $\not \!\! D_{(2)}$. 
This will be worked out explicitly below, and is related to
a fuzzy index theorem.

We note the following identities using \eq{S-yuk-id}
\bea
Tr\, \psi_{i,\a} 
\varepsilon^{\a\b}\varepsilon^{i j} (\not \!\! D_{(2)}\psi')_{j,\b}
&=& - Tr\,  
\psi_{k,\b}' \varepsilon^{\b\a}\varepsilon^{ij} (\not \!\! D_{(2)} \psi)_{j,\a} 
= Tr\, (\not \!\! D_{(2)}\psi)_{i,\a}
\varepsilon^{\a\b}\varepsilon^{i j} \psi_{j,\b}'
\label{D-symmetry}
\eea
and
\bea
Tr\, \psi_{i,\a} 
\varepsilon^{\a\b}\varepsilon^{i j} (\chi \psi)_{j,\b}'
&=& Tr\,  
\psi_{k,\b}' \varepsilon^{\b\a}\varepsilon^{ij}(\chi\psi)_{k,\a}
= - Tr\, (\chi\psi)_{i,\a}
\varepsilon^{\a\b}\varepsilon^{i j} \psi_{j,\b}'.
\label{chi-symmetry}
\eea
Therefore spinor harmonics $\psi_{i,\a}$ and $\psi'_{i,\a}$
can have a nontrivial pairing only if they have the 
same eigenvalue of 
$\not \!\! D_{(2)}$ and the opposite eigenvalue of $\chi$ 
(if applicable).
Further, observe that \eq{dirac-vanishes} amounts to
\be
Tr\, \psi_{i,\a} 
\varepsilon^{\a\b}\varepsilon^{i j} (\not \!\! D_{(2)} \psi)_{j,\b} =0,
\ee
i.e. the fuzzy Dirac operator drops out, 
and the Yukawa coupling \eq{S-yukawa} can be written as
\be
S_Y = - \frac {iN}2 \int d^4 y\,  Tr\, \psi_{i,\a} 
\varepsilon^{\a\b}\varepsilon^{i j} (\chi\psi)_{j,\b}  \, + h.c.
\label{chiralaction-fuzzy}
\ee
Therefore this model does not have the
desired 6D limit. This will be corrected below by doubling 
the fermions,
in which case the Yukawa coupling indeed induce the fuzzy Dirac operator.
But before doing that, we determine the spectrum of 
$\not \!\! D_{(2)}$.

\subsubsection{The spectrum of $\not \!\! D_{(2)}$ in the type I vacuum.} 
 
Since $\not \!\! D_{(2)}$ commutes with the $SU(2)$ group of rotations,
the eigenmodes of  $\not \!\! D_{(2)}$ in the type I vacuum \eq{vacuum-mod1}
are obtained by decomposing the spinors into irreps of $SU(2)$ 
\bea
\Psi &\in&  (2)  \otimes (N) \otimes (N) 
 = (2)\tens ((1)\oplus(3) \oplus ... \oplus (2N-1))\nn\\
&=& ((2) \oplus (4) \oplus ...  \oplus(2N))
 \, \oplus\,(\qquad \,\,\,(2) \oplus ... \oplus (2N-2))\nn\\
&=:& (\qquad \Psi_{+,(n)}  \qquad \qquad \quad\,\,\,\, 
        \oplus \qquad \Psi_{-,(n)}) .
\label{spinor-decomp}
\eea
This defines the spinor harmonics $\Psi_{\pm,(n)}$ which live in 
the $n$-dimensional representation of $SU(2)$ denoted by $(n)$ for
$n=2,4,...,2N$, excluding $\Psi_{-,(2N)}$. The eigenvalue
of $\not \!\! D_{(2)}$ acting on these states can be determined easily
using some $SU(2)$ algebra, see appendix 2:
\be 
\not \!\! D_{(2)}\Psi_{\pm,(n)} 
= E_{\delta = \pm,(n)}\,\Psi_{\pm,(n)} ,  
\label{D-spectrum}
\ee
where
\be
E_{\delta = \pm,(n)}\, \approx\, \frac{\a}2\, \left\{\begin{array}{rl} 
n, & \delta = 1,\qquad n=2,4,..., 2N \\
-n, & \delta = -1,\quad\, n=2,4,..., 2N-2
\end{array}\right.
\label{D-spectrum-explicit}
\ee
assuming $\a \approx 1$; this is exact for $\a=1$. 
We note that with the
exception of $\Psi_{+,(2N)}$, all eigenstates come in pairs
$(\Psi_{+,(n)},\Psi_{-,(n)})$ for $n=2,4,..., 2N-2$, which have opposite
eigenvalues $\pm \frac{\a}2 n$ of $\not \!\! D_{(2)}$. They are
interchanged through $\chi$,
\be
\chi \left(\begin{array}{c} \Psi_{+,(n)} \\ \Psi_{-,(n)}
\end{array}\right) =
c \left(\begin{array}{cc} 0 & 1 \\
            1 & 0 \end{array}\right)
\left(\begin{array}{c} \Psi_{+,(n)} \\ \Psi_{-,(n)}
\end{array}\right)
\label{chi-exchange}
\ee
for some $c \approx 1$, 
by virtue of the anticommutativity relation \eq{D-chi-anticomm-gen}.
 The chirality operator for the top mode
 vanishes, $\chi(\Psi_{+,(2N)}) =0$.

\subsubsection{The spectrum of $\not \!\! D_{(2)}$ in a type II vacuum.} 
 
Consider now a  type II vacuum \eq{vacuum-mod2}, 
\be
\left(\begin{array}{cc} \a_1 X_a^{N_1}\otimes \one_{n_1} & 0 \\
            0 & \a_2 X_a^{N_2}\otimes \one_{n_2}
\end{array}\right).
\ee
We decompose the spinors according to this block-structure as
\be 
\Psi_i =
\left(\begin{array}{cc} \Psi^{11}_i & \Psi^{12}_i\\ \Psi^{21}_i & \Psi^{22}_i
\end{array}\right) 
\ee 
for $i=1,2$. The analysis for
the diagonal blocks is the same as before, and they describe fermions 
in the adjoint of $SU(n_1)$ resp. $SU(n_2)$.  The off-diagonal blocks
however describe fermions in the 
bifundamental $(n_1) \times (\obar n_2)$ of $SU(n_1) \times
SU(n_2)$, and those will provide the interesting low-energy sector. 
For the moment we ignore the extra  
$SU(n_i)$ structure. Assuming $N_1 \neq N_2$,
their decomposition \eq{spinor-decomp} into irreps of
the global $SU(2)$ now reads  
\bea
\Psi^{12}_i &\in&  (2)  \otimes (N_1) \otimes (N_2) 
 = (2) \tens ((1+|N_2-N_1|)\oplus(3+|N_2-N_1|) \oplus ... \oplus (N_1+N_2-1))\nn\\
&=& \quad \qquad \qquad \qquad ((|N_2-N_1|+2) \oplus (|N_2-N_1|+4) \oplus ...  \oplus(N_1+N_2))  \nn\\
&&   \oplus ((|N_2-N_1|) \oplus (|N_2-N_1|+2) \oplus ... 
            \oplus (N_1+N_2-2))\nn\\[1ex]
&=:& (\Psi^{12}_{+,(n)}  \oplus \Psi^{12}_{-,(n)})
\label{spinor-decomp+}
\eea
defining the spinor harmonics $\Psi^{12}_{\pm,(n)}$ which live in 
the representation $(n)$ of $SU(2)$. A similar decomposition 
holds for $\Psi^{21} \in  (2)  \otimes (N_2) \otimes (N_1)$.
Then the spectrum of $\not \!\!D_{(2)}$ for 
$\Psi^{12}$ can be worked out as in appendix 2. 

For simplicity, 
we focus here on the (would-be) zero modes, which can be worked out very easily
and is the most interesting sector from the low-energy point of view. 
They are by definition the 
lowest modes  $\Psi^{12}_{-,(k)}$ in the decomposition \eq{spinor-decomp+}
of $\Psi^{12}$, where $k = |N_1-N_2|$ corresponds to the magnetic flux
induced in this type II vacuum.  It follows immediately 
from 
$(\not\!\! D_{(2)}\chi + \chi \not \!\! D_{(2)}) \Psi = O(\frac 1N)$
 that they are exact or approximate 
zero modes of $\not \!\! D_{(2)}$ (up to $O(\frac 1N)$ corrections);
this can also be checked directly.
These are precisely the 
$k$ zero modes expected from the index theorem in a monopole background
with flux $k$.

The chirality $\chi$  for the would-be zero modes
can be determined easily. 
To this end, note that they live in the subspaces 
\bea
\Psi^{12}_{-,(k)} &\in& (N+k-1) \otimes (N) \subset ((2) \otimes
(N+k)) \otimes (N), \nn\\
\Psi^{21}_{-,(k)} &\in& (N+1) \otimes (N+k)\subset ((2) \otimes (N))
\otimes (N+k).
\eea
Note that this involves the ``anti-parallel''
resp. ``parallel'' tensor product for the first 2 factors.
Therefore 
$\Phi = \sigma^a \phi_a + \phi_0 \one \approx -\frac N2$
if acting from the left on  $\Psi^{12}_{-,(k)}$ , and  
$\Phi\approx +\frac N2$ if acting on $\Psi^{12}_{-,(k)}$. 
Since $\frac 1N\,\Phi^L$ agrees with $\chi$ up to $1/N$, it follows that
\bea
\chi(\Psi^{12}_{-,(k)}) &=& c^{12}\, \Psi^{12}_{-,(k)},
\qquad \qquad c^{12} \approx - 1,  \nn\\
\chi(\Psi^{21}_{-,(k)}) &=& c^{21}\, \Psi^{21}_{-,(k)}, 
\qquad\qquad c^{21} \approx  1.
\label{zero-modes-chiral}
\eea
The chirality can be computed more generally 
using \eq{chi-explicit}.

\subsection{Doubling the fermions}
\label{sec:double-fermions}

The fact that we did not arrive at the expected 6-dimensional description
of fermions in the previous section can be understood as follows: 
Usually, in order to introduce fermions in 4+2 resp. 4+3
 dimensions one starts with 
the 6- resp. 7-dimensional Clifford algebra
\be
\Gamma^A = (\Gamma^{\mu},\Gamma^a) = (\one \otimes \gamma^\mu,
\sigma^a \otimes i\gamma_5)
\label{gamma-6D}
\ee
which act on $\C^2 \otimes \C^4$  and satisfies
$(\Gamma^A)^\dagger = \eta^{AB} \Gamma^B$ where 
$\eta^{AB} = (1,-1,...,-1)$ is the 7-dimensional Minkowski metric.
This corresponds to 4-dimensional Dirac fermions tensored with
2-dimensional Dirac fermions. The 
6D chirality operator is given by
\be
\Gamma =  \gamma_5 \chi ,
\label{chiral-6D}
\ee
where $\chi = -i\sigma_1 \sigma_2 = \sigma_3$. 
In particular, 6-dimensional chiral fermions necessarily contain both
chiralities from
the 4D point of view. In order to reproduce this in our model, 
we should therefore start with 4D Dirac fermions, 
i.e. double the Weyl fermions introduced above. 
Hence consider a doublet of Weyl fermions\footnote{It is easy to check
that the gauge sector remains to be asymptotically free.}
$\psi_{i,r;\a}(y)$ for $r\in\{1,-1\}$ in the adjoint of $SU(\cN)$,
\be
\Psi = \left(\begin{array}{c} \psi_{i,1;\a} \\ 
    \psi_{i,2;\a} \end{array}\right)
\equiv \left(\begin{array}{c} \rho_{i,\a} \\ 
       \eta_{i,\a} \end{array}\right) .
\label{etarho}
\ee
They transform as a doublet under 
$SU(2)_R$ acting on the $r\in\{1,-1\}$ indices, which 
may or may not be a symmetry of the action. 
This  $SU(2)_R$
contains in particular the $U(1)_R$ symmetry
\be
\left(\begin{array}{c} \rho_{i,\a} \\ 
       \eta_{i,\a} \end{array}\right)
\rightarrow\,  e^{i a R}\,\left(\begin{array}{c} \rho_{i,\a} \\ 
       \eta_{i,\a} \end{array}\right)
=  \left(\begin{array}{c} e^{i a}\, \rho_{i,\a} \\ 
      e^{-i a}\, \eta_{i,\a} \end{array}\right)
\label{etarho-vector-U1}
\ee
which prevents self-couplings. This will later be identified as 
``vector'' $U(1)$ charge, with generator
\be
R = \left(\begin{array}{cc} 1 & 0 \\ 
       0 & -1 \end{array}\right) .
\label{R-def}
\ee
The only non-vanishing mass term  is 
\be
S_m = \int d^4 y\,  Tr\, m'\, \psi_{i,r,\a} 
\varepsilon^{\a\b} \varepsilon^{i j} \varepsilon^{rs} \psi_{j,s,\b}  + h.c. 
\label{majorana-mass-2W}
\ee
since any symmetric combination vanishes. 
Here $m'$ might be complex, which will be important below.
Note that this term is automatically 
invariant under the global $SU(2)_R$.

Now consider the Yukawa couplings.
Using \eq{S-yuk-id} and the above definitions, 
the most general Yukawa interaction can be written as 
\bea
S_Y &=&  \int d^4 y\,  Tr\, \Big( \psi_{i,r,\a} 
\varepsilon^{\a\b}\varepsilon^{i j} \varepsilon^{rs} 
\big(e (\not \!\! D_{(2)}-1)\psi + f  R\chi\psi \,\big)_{k,s,\b}\,  \nn\\
&& \qquad \qquad\quad + \big( h_1 \rho_{i,\a} (\chi\rho)_{j,\b} + h_2
\eta_{i,\a} (\chi\eta)_{j,\b} \big) \varepsilon^{\a\b}\varepsilon^{i j}\Big)
  \quad  + \, h.c. 
\label{S-yukawa-2W}
\eea
for constants $e,f,h_i$. Imposing the ``vector'' $U(1)_R$
invariance implies $h_1 = h_2 = 0$, while
imposing the full $SU(2)_R$ symmetry implies $f=h_i=0$,
leaving the shifted Dirac operator $(\not \!\! D_{(2)} -1\,)$ 
on the internal sphere as only possible Yukawa interaction. 
We will first consider the $SU(2)_R$ - symmetric case, and postpone the 
general case to section \ref{sec:su2-breaking}. 
Redefining the mass parameter $m' = m + e$, we thus
obtain\footnote{more precisely $m' = m + \frac eR$; 
recall that we dropped the radius parameter $R$}
\be
S_Y + S_m =   \int d^4 y\,  Tr\,  \psi_{i,r,\a} 
\varepsilon^{\a\b}\varepsilon^{i j} \varepsilon^{rs} 
\big(e\not \!\! D_{(2)}\psi + m\psi\big)_{k,s,\b} \quad  + \,h.c. 
\label{S-yukawa-2W-symm}
\ee 
Including $\psi_0$ in the extended formalism of section
\ref{sec:SU2N} naturally
suggests that $m=0$, but we cannot strictly rule out a
bare mass term at this point.

\paragraph{Kinetic term}

The kinetic term of the action is as in \eq{kinetic-fermions},
\be
S_K = \int d^4 y\, Tr\, \Psi^\dagger 
i \sigma^\mu (\partial_\mu + [A_\mu,.]) \Psi
 =  \int d^4 y\, Tr\, (\psi_{i,r,\a})^\dagger 
i ({\obar\sigma}^\mu)^{\dot\a\b} (\partial_\mu + [A_\mu,.])
\psi_{i,r,\b} 
\label{kinetic-term}
\ee
which is invariant under all the symmetries. We will now show that the combined
action $S_K + S_Y + S_m$ is naturally interpreted as a 6D 
action for a Dirac fermion on $M^4 \times S^2$, which at low energy
behaves like a compactified 4D action on $M^4$.

\subsubsection{Effective 6D Dirac fermion}

We can combine the $r=1,2$  components of the 4 Weyl fermions
\eq{etarho} into a Dirac fermion,
\be
\Psi_{D} = \left(\begin{array}{c} \psi_{i,1;\a} \\ 
     \varepsilon_{i j}\;\varepsilon^{\dot\a\dot\b}\; (\psi_{j,2;\b})^{\dagger} \end{array}\right)
\equiv \left(\begin{array}{c} \rho_{i,\a} \\ 
       \obar\eta_i^{\dot\a} \end{array}\right)
 \equiv \left(\begin{array}{c} \rho_{\a} \\ 
       \obar\eta^{\dot\a} \end{array}\right)
\quad \in \C^4 \otimes \C^2 \otimes Mat(\cN,\C).
\label{Dirac-spinor}
\ee
Then the kinetic term can be written as
\bea
S_K &=& \int d^4 y\, Tr\, \obar\Psi_D 
i \gamma^\mu (\partial_\mu + [A_\mu,.] ) \Psi_D  \nn\\
 &=& \int d^4 y\, Tr\, \Big(
\rho_{\a}^{\dagger}\,i(\obar{\sigma}^\mu)^{\dot\a\b}(\partial_\mu + [A_\mu,.]) 
\rho_{\b}
+ (\obar\eta^{\dot\a})^{\dagger}\, i(\sigma^{\mu})_{\a\dot\b}(\partial_\mu + [A_\mu,.])
\obar\eta^{\dot\b} \Big) , 
\label{S-dirac-full}
\eea
where 
\be
\gamma^\mu = \(\begin{array}{cc}0 & \sigma^\mu \\
                               \obar\sigma^\mu & 0 \end{array} \)
\ee
acts as usual on the Dirac spinors 
\eq{Dirac-spinor}.
The hermitian extensions of the mass term \eq{majorana-mass-2W}
with complex mass can be written as
\be
S_m = \int d^4 y\, Tr\,
    (\tilde m - i m') \rho_{\a}^{\dagger}\obar\eta^{\dot\a} 
 + (\tilde m +i m') {\obar\eta^{\dot\a}}^{\dagger}\rho_{\a}
 = \int d^4 y\, Tr\, \obar\Psi_D (\tilde m + i\gamma^5 m'\,) \Psi_D .
\ee
Note that $\int d^4 y\, Tr\, \obar\Psi_D i\gamma^5 \Psi_D $
will become part of the Dirac operator. 
Furthermore, using \eq{D-symmetry} we have 
\bea
\int d^4 y\, Tr\,
\psi_{2,j\b} \varepsilon^{ji} \varepsilon^{\b\a}  i\not \!\! D_{(2)} 
\psi_{1,i\a}
 &=&\int d^4 y\, Tr\,(\obar\eta^{\dot\a})^{\dagger }
 i\not \!\! D_{(2)} \rho_{\a} \nn\\
 &=& - \int d^4 y\, Tr\,
\psi_{1,j\b} \varepsilon^{ji} \varepsilon^{\b\a}  i\not \!\! D_{(2)} 
\psi_{2,j\a} ,
\eea 
 and the 
Yukawa couplings \eq{S-yukawa-2W} in the $SU(2)_R$ -symmetric case
can  be written as\footnote{we can assume
  that $e$ is real by rotating the phases of the fermions if necessary.} 
\be
S_Y = e \int d^4 y\, Tr\, \obar\Psi_D 
  i\gamma_5 ( \not \!\! D_{(2)} -1) \Psi_D
= e\int d^4 y\, Tr\, \left(
\rho_{\a}^{\dagger } i (\not \!\! D_{(2)} -1)
\obar\eta^{\dot\a}
- (\obar\eta^{\dot\a})^\dagger  i (\not \!\! D_{(2)} -1) \rho_{\a} \right).
\label{S-dirac-yuk}
\ee
The $SU(2)_R$ symmetry is now hidden but still 
holds due to Grassmann antisymmetry. 
In particular, note that 
$\gamma^5 = \(\begin{array}{cc}1 & 0\\ 0 & -1 \end{array}\) = R$ 
 \eq{R-def}  ensures hermiticity, since
\bea
\(\int d^4 y\, Tr\, 
\rho_{\a}^{\dagger }  i\not \!\! D_{(2)} 
\obar\eta^{\dot\a}\)^\dagger 
&=& -\int d^4 y\, Tr\, (\obar\eta^{\dot\a})^{\dagger }
 i\not \!\! D_{(2)} \rho_{\a}.
\eea
Writing again $m' = m +e$, we obtain
\bea
S_K + S_Y + S_m = S_{6D} 
&=& \int d^4 y\,  Tr\,\obar\Psi_D \left(
i \gamma^\mu (\partial_\mu + [A_\mu,.]) 
+ e  i\gamma_5 \not \!\!D_{(2)} + \tilde m +i \gamma_5 m\right)\Psi_D  \nn\\
&\equiv& \int d^4 y\,  Tr\,\obar\Psi_D (\not \!\!D_{(6)} 
+ \tilde m +  i \gamma_5  m)\Psi_D .
\label{fermion-action-6D}
\eea
Thus apart from the 2 distinct mass parameters,
the fermionic 
action \eq{fermion-action-6D} can indeed be interpreted
as gauged Dirac operator on $M^4 \times S^2_N$. 
Note again that this is a result, which was not 
imposed on the model in any way. 
We remark that the bare Dirac mass  is expected to run only
weakly, being protected by the (approximate) 6D chiral symmetry.

\subsubsection{Fermionic low-energy action and Kaluza-Klein modes}

To obtain the appropriate low-energy action in 4 dimension, we 
should organize the fermions in terms of the 
eigenmodes \eq{D-spectrum} $\psi_{\pm,(n)}$
of $\not \!\! D_{(2)}$, 
i.e. $\not \!\! D_{(2)} \psi_{\pm,(n)} = E_{n,\pm} \psi_{\pm,(n)}$.
Consider first a type I vacuum. Then 
\be
\int d^4 y\,  Tr\,\obar\Psi_D  i\gamma_5 \not \!\!D_{(2)}\Psi_D
= \int d^4 y\, Tr\, \sum_{n,\pm} i E_{n,\pm} \left(
\rho_{\pm,(n),\a}^{\dagger}  
\obar\eta^{\dot\a}_{\pm,(n)}
- (\obar\eta_{\pm,(n)}^{\dot\a})^\dagger  \rho_{\pm,(n),\a} \right) 
\label{S-dirac-yuk-n}
\ee
using \eq{S-dirac-yuk},
\be
(\obar\eta_{\pm,(n)}^{\dot\a})^{\dagger i} = \eta_{\pm,(n);j\b}\,
\varepsilon^{ji}\;\varepsilon^{\b\a} \nn ,
\ee
and the orthogonality of the eigenstates, which follows from
\be
\int d^4 y\, Tr\, (\not \!\! D_{(2)} \obar\eta^{\dot\a})^\dagger
\rho_{\a}
= \int d^4 y\, Tr\, (\obar\eta^{\dot\a})^{\dagger }
 \not \!\! D_{(2)} \rho_{\a} 
\ee
as well as $\int d^4 y\, Tr\, (\chi(\obar\eta^{\dot\a}))^\dagger
\rho_{\a}
= \int d^4 y\, Tr\, (\obar\eta^{\dot\a})^{\dagger }
 \chi(\rho_{\a})$.
Hence the fermions  naturally pair up into 4D Dirac fermions as follows
\be
\Psi_{+,D,(n)} = \left(\begin{array}{c} \rho_{+,(n),\a} \\ 
              i\, \obar\eta_{+,(n)}^{\dot\a} \end{array}\right),
\qquad \Psi_{-,D,(n)} = \left(\begin{array}{c}  \rho_{-,(n),\a} \\ 
            i\,  \obar\eta_{-,(n)}^{\dot\a} \end{array}\right),
\label{Dirac-spinor-KK}
\ee
and we obtain
\be
S_Y + S_m = e\int d^4 y\, Tr\, \sum_{n}   \left(
   (E_{n,+}+m)\,\obar\Psi_{+,D,(n)} \Psi_{+,D,(n)}  
+ (E_{n,-}+m)\,\obar\Psi_{-,D,(n)} \Psi_{-,D,(n)} \right)
\label{S-dirac-yuk-n-2}
\ee 
dropping $\tilde m$ for simplicity\footnote{$\tilde m$ would lead to
  an additional  shift in the KK mass spectrum}.
The sign of $(E_{n,\pm}+m)$ is irrelevant and can be absorbed by a
phase rotation of the negative eigenmodes.
The kinetic term which can be written as
\be
\int d^4 y\, Tr\, \sum_{n} \big(\obar\Psi_{+,D,(n)}
i \gamma^\mu (\partial_\mu + g[A_\mu,.] ) \Psi_{+,D,(n)} 
+ \obar\Psi_{-,D,(n)}
i \gamma^\mu (\partial_\mu + g[A_\mu,.] ) \Psi_{-,D,(n)}\big) ,
\ee
and we obtain the expected KK tower of massive 4D Dirac fermions with
masses 
\be
m_{\pm,D,n} =  e |E_{n,\pm} + m| \neq 0 
\label{Dirac-masses}
\ee
which are non-zero unless $m$ is adjusted very particularly. 
In particular, there are no massless 
fermions even if the bare mass
$m=0$, since
$E_{n,\pm} \neq 0$ in the type I vacuum.
This will change in the type II vacuum, as discussed in 
section \ref{sec:type2-zeromodes}.
In particular note that also the top modes 
\be
\rho_{+,(2N)}, \quad \eta_{+,(2N)}
\ee
form very massive Dirac fermions in 4D in the non-chiral case,
and play no role at low energies.
Their role in the chiral case is more subtle and discussed below. 

Another comment is in order.
We recall from \cite{Aschieri:2006uw} that the effective radius of the
internal 2-sphere is given by $r_{S^2} = \frac {\a}g\, R$, where $g$ is
the gauge coupling. According to \eq{Dirac-masses} and
\eq{D-spectrum-explicit}, the fermions see
the effective radius 
\be
\tilde r_{S^2} = \frac {\a}e\, R, 
\label{radius-fremions}
\ee
which differs
in general from $r_{S^2}$ and depends on the Yukawa coupling $e$. 
This shows that the present framework provides in fact a slight generalization
of the conventional compactification, in accord with the 2 mass terms 
found in \eq{fermion-action-6D}.

\subsection{Chirality and Kaluza-Klein  modes}
\label{subsec:chirality}

We now want to impose a chirality constraint 
on $M^4 \times S^2_N$. The first attempt might be 
to impose $\Gamma \Psi = \Psi$ using  the 6D chirality operator
$\Gamma$ \eq{chiral-6D}. 
This is however not sensible because 
$\Gamma^2 \neq 1$. A consistent 6D chirality constraint could be
\be
\tilde\Gamma \Psi = \Psi ,
\ee
where 
\be
\tilde\Gamma = \gamma_5 \tilde \chi, \qquad \quad
\tilde\chi = \chi_+ - \chi_-
\ee
and $\chi_\pm$
denotes the spectral
projectors on the positive and negative eigenvectors of 
$\chi$. Then we can consider chiral 6D fermions which satisfy
\be
\tilde\chi \rho_{\a} =  \rho_\a, \qquad 
\tilde\chi\obar\eta^{\dot\a} = - \obar\eta^{\dot\a} . 
\ee
It is shown in appendix 2 that
$\chi(\obar\eta^{\dot\a}) = - \obar{\chi\eta_\a}$, hence
this is equivalent to 
\be
\tilde\chi\rho_{\a} = \rho_{\a}, \qquad
\tilde \chi\eta_{\a} = \eta_{\a} ,
\label{chiral-explicit}
\ee
i.e. the same conditions apply to $\rho$ and $\eta$.
The (would-be) zero modes of the type II vacuum
and the top modes 
have multiplicity one, and are therefore 
either admitted or dismissed by this chirality 
constraint\footnote{actually the top modes are discarded
since $\chi$ vanishes}. 
For the other eigenmodes of $\not \!\!D_{(2)}$,  the chirality
operator $\chi$ exchanges the positive and the negative 
eigenmodes \eq{chi-exchange},
\be
\chi \rho_{+,(n),\a} = c \rho_{-,(n),\a}, \qquad
\chi \eta_{+,(n),\a} = c \eta_{-,(n),\a}  
\ee
for some $c \neq 0$, or equivalently 
\be
\chi \obar\rho_{+,(n)}^{\dot\a} = -c\obar\rho_{-,(n)}^{\dot\a}, \qquad
\chi \obar\eta_{+,(n)}^{\dot\a} = -c\obar\eta_{-,(n)}^{\dot\a}.
\ee
This reduces the degrees of freedom by half, and
\be
\Psi_{+,D,(n)} = \left(\begin{array}{c} \rho_{+,(n),\a} \\ 
              i\, \obar\eta_{+,(n)}^{\dot\a} \end{array}\right),
\qquad c\Psi_{-,D,(n)} = \left(\begin{array}{c}  \chi\rho_{+,(n),\a} \\ 
            i\,  \chi\obar\eta_{+,(n)}^{\dot\a} \end{array}\right)
 = \chi \gamma_5\Psi_{+,D,(n)} .
\label{Dirac-spinor-KK-chiral}
\ee
Then the contribution of $\Psi_{-,D,(n)}$ in \eq{S-dirac-yuk-n-2} 
coincides with the 
one from $\Psi_{+,D,(n)}$, 
leading to a single multiplet of 4D massive Dirac fermions 
$\Psi_{+,D,(n)}$. 
Therefore in the 6D chiral case
there is a single multiplet of 4D massive Dirac fermions 
$\Psi_{+,D,(n)}$ with mass $E_{n,+}$, as opposed to 
2 multiplets in the non-chiral case.
However, there are  no massless fermions in the type I vacuum.

\paragraph{Alternative chirality operator}
There is an alternative  possibility
to define a chirality operator on the fuzzy sphere, 
which is related (but not identical) 
to the Ginsparg-Wilson approach of \cite{Balachandran:2003ay}.
The basic observation is the following: consider the (antihermitian) 
$2\cN \times 2\cN$ matrix
\be
\Phi = \phi_0 \one_2 + \phi_a \sigma_a
\ee
with $\phi_0 = -\frac {i}2$ as in \eq{Phi-extended}.
It satisfies  $\Phi^2 \approx c\,\one$ in and near any of the vacua 
of interest, and we assume that
$\Phi$ has no zero eigenvalue. 
We can thus define 
\be
\tilde \Phi:=\frac{i\Phi}{|i\Phi|}.
\label{tildephi-def}
\ee 
Then 
\be
\tilde \chi' \Psi := \tilde \Phi \Psi 
\label{chi-tildeprime--def}
\ee  
is a good chirality operator for 
the fuzzy sphere, and
we can  impose the alternative 6D chirality constraint
\be
\tilde\Gamma' \Psi =  \gamma_5 \tilde \chi'\Psi = \Psi .
\label{chirality-constraint-prime}
\ee
This is particularly natural if $\Phi^2 \equiv c\,\one$, 
which is an interesting constraint studied in the  
$SU(2\cN)$- extended formalism of section \ref{sec:SU2N}. 
It is easy to see using \eq{Phi-chi-D-relation-1} 
that $\tilde \chi'$  agrees with $\tilde\chi$ 
on the low-energy modes.
However, $\tilde \chi$ leads to a problem with the top modes:

\paragraph{Top modes}

Consider the top modes 
\be
\rho_{+,(2N)}, \quad \eta_{+,(2N)}
\ee
in the chiral case. Their 2D chirality $\chi$ vanishes identically,
\be
\chi\rho_{+,(2N)}=0,
\ee
which follows from
$(\not \!\! D_{(2)}\chi + \chi \not \!\! D_{(2)}) \Psi = O(\frac 1N)$
combined with the fact that $\not \!\! D_{(2)} = O(N)$; 
this can also be computed directly. 
Therefore the chiral projection $\tilde\Gamma \Psi = \Psi$
removes these top modes.

On the other hand, the alternative 6D chirality
constraint \eq{chirality-constraint-prime} gives
$\tilde\chi'\rho_{+,(2N)} = \rho_{+,(2N)}$ and the
same for $\eta$. This can easily be seen by noting that
 the top modes only contain the maximal spin as seen by the
intertwiner $\Phi$. 
Therefore the constraint $\tilde\Gamma' \Psi = \Psi$
preserves $\rho_{+,(2N),\a}$ but 
excludes $\obar\eta_{+,(2N)}^{\dot\a}$ and hence $\eta_{+,(2N),\a}$. 
The surviving $\rho_{-,(2N)}$ cannot acquire any mass
and hence form a large massless 
multiplet; indeed any self-coupling term 
\be
Tr \rho_{+,(2N),i \a} \varepsilon^{i j}\;\varepsilon^{\a\b}\;
\rho_{+,(2N),j \b}
\ee
vanishes identically. Since it does couple to the 
gauge field, such a large massless multiplet is not acceptable.
This is a serious problem with the chirality projector
$\tilde\Gamma'$, which might be overcome by adding a second ``mirror''
copy of fermions.

\paragraph{6D Majorana condition}

A fuzzy analog of the 
 6D Majorana condition $\Psi_D^* = C \Psi_D$ amounts in the
component form  to
\be
\rho_{i,\a} = \eta_{i,\a} .
\ee
This leads to the minimal approach 
of section \ref{sec:Weylfermions},
which did not give the desired 6D interpretation.
The reason appears to be again that the 
fuzzy sphere does see some trace of the embedding 3rd dimension, and
therefore 
does not seem to allow a Majorana condition.

\subsection{Type II vacuum: zero modes and chirality}
\label{sec:type2-zeromodes}

In a vacuum of type \eq{vacuum-mod2}, we  decompose the spinors as
\be \Psi_\a =
\left(\begin{array}{cc} \Psi^{11}_\a & \Psi^{12}_\a\\ \Psi^{21}_\a & \Psi^{22}_\a
\end{array}\right) 
\ee 
according to \eq{vacuum-mod2}. The analysis for
the diagonal blocks $\Psi^{11},\Psi^{22}$ is the same as before. 
In particular, there are no zero modes even if the 
bare mass term $m$ vanishes. 
 
The off-diagonal blocks
however describe fermions in the bifundamental of $SU(n_1) \times
SU(n_2)$, more precisely $\Psi^{12}$ 
lives in $(n_1) \times (\obar n_2)$ of $SU(n_1) \times SU(n_2)$, 
while $\Psi^{21}$ lives in $(\obar n_1) \times (n_2)$.
Their arrangement into Dirac fermions is as follows
\be
\Psi_{D}^{12} = \left(\begin{array}{c} \psi_{i,1;\a}^{12} \\ 
     \varepsilon_{i j}\;\varepsilon^{\dot\a\dot\b}\; (\psi_{j,2;\b}^{21})^{\dagger} \end{array}\right)
\equiv \left(\begin{array}{c} \rho_{i,\a}^{12} \\ 
       (\obar\eta_i^{\dot\a})^{12} \end{array}\right)
 \equiv \left(\begin{array}{c} \rho_{\a}^{12} \\ 
       (\obar\eta^{\dot\a})^{12} \end{array}\right)
\label{Dirac-spinor-II}
\ee
and similarly for $\Psi_{D}^{21}$,
noting that the block index $12$ resp. $21$ gets interchanged by the
conjugation. The 4D chirality $\gamma_5$ is now manifest.
The contribution of these off-diagonal blocks to the Yukawa can be
written as
\be
S_Y = e\int d^4 y\, Tr\, \sum_{n >0}   \left(
   E_{n,+}\,(\obar\Psi^{21}_{+,D,(n)} \Psi^{12}_{+,D,(n)}  
   + \obar\Psi^{12}_{+,D,(n)} \Psi^{21}_{+,D,(n)})
+ E_{n,-}\,(...) \right).
\label{S-dirac-yuk-n-vac2}
\ee 
This is hermitian,
\be
\int d^4 y\, Tr\, (\obar\Psi_D^{21} \Psi^{12}_D)^{\dagger} = 
\int d^4 y\, Tr\, (\Psi^{12}_D)^\dagger \gamma_0^\dagger \Psi^{21}_D = 
\int d^4 y\, Tr\, \obar\Psi^{12}_D \Psi^{21}_D 
\ee
In particular, \eq{S-dirac-yuk-n} gives
\bea 
S_Y &=& -e\int d^4 y\, Tr\, \sum_{n,\pm} i E_{n,\pm} \left(
 (\obar\eta_{\pm,(n)}^{12;\dot\a})^\dagger  \rho_{\pm,(n),\a}^{21} 
 + (\obar\eta_{\pm,(n)}^{21;\dot\a})^\dagger  \rho_{\pm,(n),\a}^{12} \right)
 + h.c.  \nn\\
&=& -e\int d^4 y\, Tr\, \sum_{n,\pm} i E_{n,\pm} \left(
\eta_{\pm,(n),i,\a}^{12} \rho_{\pm,(n),,j\b}^{21}
 -\rho_{\pm,(n),i,\a}^{12} \eta_{\pm,(n),j,\b}^{21} \right)
\varepsilon^{\a\b}\varepsilon^{ij}
 + h.c. 
\label{S-dirac-yuk-n-3}
\eea
This makes explicit the $SU(2)_R$ symmetry acting on $(\rho,\eta)$.
Recall that without this doubling, all diagonal 
KK modes would be massless.

We now want to understand the low-energy sector of the 
model.

\paragraph{Would-be zero modes: non-chiral case}

Let us focus on the (would-be) zero modes 
$\Psi_{-,D,(k)}^{12}, \Psi_{-,D,(k)}^{21}$ for $k=N_1-N_2$, 
which determine the low-energy physics. The $\Psi_{-,D,(k)}^{12}$ 
provide $k$ families of fermions
transforming
in $(n_1) \otimes (\obar n_2)$ of $SU(n_1) \times SU(n_2)$. 
Similarly, the (would-be) zero modes 
$\Psi_{-,D,(k)}^{21}$ provides  $k$ fermions
transforming
in $(\obar n_1) \otimes (n_2)$ of $SU(n_1) \times SU(n_2)$.
In the non-chiral case, they are massless only if the bare mass
vanishes. 

Consider the chirality of these modes. 
We have seen in \eq{zero-modes-chiral}
that they are eigenmodes of the 2D chirality operator 
with 
\be
\tilde\chi(\Psi^{12}_{-,D,(k)}) = -\Psi^{12}_{-,D,(k)},  \qquad
\tilde\chi(\Psi^{21}_{-,D,(k)}) = \Psi^{21}_{-,D,(k)},
\ee
and
it follows as usual that they are exact or approximate 
zero modes of $\not \!\!D_{(2)}$. This 
is due to the monopole flux with strength $k$ on the fuzzy sphere.
We thus get 2 (almost-) massless Dirac fermions
$\Psi^{12}_{-,D,(k)}$ and $\Psi^{21}_{-,D,(k)}$. However, despite  their
special role there is nothing in the  non-chiral theory
which prevents them from acquiring a mass term of the form
$\int Tr \obar \Psi^{12}_{-,D,(k)} \Psi^{21}_{-,D,(k)}$. 
These terms are
explicitly present in \eq{S-dirac-yuk-n-3}, 
and they vanish only if
the bare mass  vanishes. While this is natural in the 
extended $SU(2\cN)$ formalism of section \ref{sec:SU2N}, it is not
forced by any symmetry and therefore  
amounts to some fine-tuning. This is why we call
them ``would-be zero modes''.

Moreover, even if these would-be zero modes are exactly massless, the
low-energy theory is not complex, since every fermion in 
$(n_1) \otimes (\obar n_2)$ has a counterpart in the conjugate 
representation $(n_2) \otimes (\obar n_1)$. We therefore find 
essentially ``mirror fermions'', 
which will be discussed below.

\paragraph{Zero modes: chiral case}

Imposing the 6D chirality constraint $\tilde\Gamma \Psi_D = -\Psi_D$ implies
using \eq{chiral-explicit} and \eq{zero-modes-chiral}
that $\Psi_{-,D,(k)}^{21}$ i.e. 
$\rho_{-,(k),\a}^{21}$ and $\eta_{-,(k),\a}^{21}$ are
discarded, since they have the wrong  chirality. 
Then only $\rho_{-,(k),\a}^{12}$
and  $\eta_{-,(k),\a}^{12}$
survive which both live in $(n_1) \times (\obar n_2)$
(or equivalently $\obar\rho_{-,(k)}^{\dot\a 21}$ and $\obar\eta_{-,(k)}^{\dot\a 21}$
which live in $(n_2) \times (\obar n_1)$), and
there is no way to write down a mass term for these modes. Hence
we have a doublet of {\em exactly massless chiral fermions}.
This is the desired mechanism based on the index theorem.
The  reason for the doubling encountered here 
is that we started with Dirac fermions in the adjoint of $SU(\cN)$.

In the case of unbroken gauge group $SU(3) \times SU(2)$, these
zero modes 
could be interpreted as $k$ left-handed quarks $(u,d)_L$. 
The $U(1)$ generator is  
$\propto (\frac 1{3N_1},\frac 1{3N_1},\frac 1{3N_1},-\frac 1{2N_2},-\frac
1{2N_2})$, which for large $N_1 \approx N_2$ is close to but
different from the standard unbroken $U(1)$ as obtained e.g. from 
the $SU(5)$ GUT. By extending\footnote{this is particularly natural 
from the point of view of noncommutative field theory} the basic 
gauge group to $U(\cN)$, 
one could obtain the
more standard $U(1)$ generator
$(\frac 1{3},\frac 1{3},\frac 1{3},-\frac 1{2},-\frac
1{2})$.
Further prospects to obtain realistic models are
discussed in section \ref{sec:fluxons-ssb}.

\paragraph{Top modes}

Consider now the top modes
\be
\rho_{-,(N_1+N_2)}, \quad \eta_{-,(N_1+N_2)}
\ee
for the type II vacuum. 
In the non-chiral case, these top modes form very massive
Dirac fermions in 4D which play no role at low energies.

In the chiral case, the 
same remarks as for the type I vacuum apply to the diagonal 
blocks, leading to large massless multiplets in the case of the
$\tilde\Gamma'$ projector.   
The off-diagonal top modes could acquire a mass even using the 
$\tilde\Gamma'$
projector, since $\rho_{-,(N_1+N_2),i \a}^{12}$ and 
$\rho_{-,(N_1+N_2),i \a}^{21}$ have the same 
$\tilde\Gamma'$ -chirality, and
\be
Tr\, \rho_{-,(N_1+N_2),i \a}^{12} \varepsilon^{i j}\;\varepsilon^{\a\b}\;
\rho_{-,(N_1+N_2),j \b}^{21} \,
\neq 0
\ee
is non-vanishing. Therefore only the diagonal ones remain to be
problematic. For the $\tilde\Gamma$ -chirality,
all top modes are projected out.

\subsection{Breaking $SU(2)_R$}
\label{sec:su2-breaking}

One of the motivations for the
global $SU(2)_R$ symmetry is the fact that
the only possible mass term \eq{majorana-mass-2W} is
automatically invariant under $SU(2)_R$.
However, the Yukawa couplings \eq{S-yukawa-2W} do not necessarily
preserve it. Let us now consider the case where the Yukawa couplings
break $SU(2)_R$. This is interesting in particular because then 
the doubling of the fermions becomes asymmetrical, i.e.
the ``mirror'' fermions may have different low-energy
properties. This is of course essential from 
a phenomenological point of view.

The most general Yukawa coupling was given in \eq{S-yukawa-2W}. 
Using 
\bea
Tr\,\rho_{i,\a} \chi(\eta_{j,\b}) \varepsilon^{\a\b}\varepsilon^{i j} 
&=& Tr\eta_{i,\a}\chi(\rho_{j,\b})\varepsilon^{\a\b}\varepsilon^{i j},\nn\\
Tr\,\rho_{i,\a} \eta_{j,\b} \varepsilon^{\a\b}\varepsilon^{i j}
&=& - Tr\eta_{i,\a}\rho_{j,\b}\varepsilon^{\a\b}\varepsilon^{i j}
\eea
it simplifies as 
\bea
S_Y &=& \int d^4 y\,  Tr\,  \psi_{i,r,\a} 
\varepsilon^{\a\b}\varepsilon^{i j} \varepsilon^{rs} 
(\not \!\! D_{(2)} -1) \psi_{k,s,\b} \nn\\
 &&\quad + \varepsilon^{\a\b}\varepsilon^{i j}
\Big(2 f \rho_{i,\a} \chi(\eta_{j,\b})  
+ h_1 \rho_{i,\a} \chi(\rho_{j,\b}) 
+ h_2 \eta_{i,\a} \chi(\eta_{j,\b})\Big) .
\label{S-yukawa-3}
\eea
Together with the mass term, this gives 
the general mass matrix for the (would-be)
zero modes of $\not \!\! D_{(2)}$
in a type II vacuum
\be
\left(\begin{array}{cc} h_1 & m+f\chi\\ m-f\chi & h_2
\end{array}\right)
\ee
acting on $\left(\begin{array}{c} \rho \\ \eta \end{array}\right)$.
For illustration, consider 
the case $h_1=h_2=0$. One can  then write the additional term as
6D pseudo-scalar 
\be
\int d^4 y\, Tr f\, \obar\Psi\, i\Gamma \,\Psi .
\ee
For the would-be zero-modes in a type II vacuum, 
this gives explicitly the additional terms
\be
\int d^4 y\,  Tr\,  
\Big((m + f)\, \rho_{i,\a}^{12} \eta_{j,\b}^{21} + (m-f)\,\rho_{i,\a}^{21} \eta_{j,\b}^{12} \Big) \varepsilon^{\a\b}\varepsilon^{i j}
\ee
If e.g. $m=\pm f$, then $(\rho^{12}, \eta^{21})$ 
form a massive Dirac fermion, while $(\rho^{21}, \eta^{12})$ 
remain massless and form a single mirror pair of fermions with 
opposite chirality. Therefore this removes\footnote{at the expense of some fine-tuning} the
fermion doubling, but it does not remove the fact that each
fermion has a mirror partner with opposite chirality 
and opposite quantum numbers.

\section{Extended model and $SU(2\cN)$ structure}
\label{sec:SU2N}

In this section we point out that the degrees of freedom
of this model are naturally arranged in $SU(2\cN)$ 
representations. For the fermions this is elaborated 
in section \eq{fermions-extended}.
This gives a natural relation with the twisted
picture discussed in \cite{Andrews:2005cv}. 
Similarly, the scalars $\phi_i$
are naturally arranged
as $\Phi = \phi_0 + \phi_a \sigma_a$ by adding
a further component $\phi_0$; this has been 
anticipated several times. 
There are 2 motivations for this point of view: first, it naturally
leads to the correct constant shift in the fuzzy Dirac
operator \eq{fuzzydirac-0}, see \eq{S-yukawa-2} below; however this is
only suggestive.
The main motivation is that it suggests a 
$SU(2\cN)$-invariant constraint
$\Phi^2 \propto \one$, which is known to provide an alternative
description of Yang-Mills 
theory on $S^2_N$ \cite{Steinacker:2007iq}. 
This in turn is related to the alternative
definition of a chirality projection \eq{chirality-constraint-prime},
 \eq{PhiPsi-constraint}.
While we are not able at present to show that it 
is consistent at the quantum level, this appears to be the 
best candidate for a chirality constraint. 
All this points to an underlying $SU(2\cN)$ structure, which
certainly deserves further investigation, 
including possible SUSY versions.

\subsection{$SU(2\cN)$ structure and Yukawa 
coupling}
\label{fermions-extended}

It is natural to collect the fermions into a $2\cN \times 2\cN$ matrix
as follows
\be
\Psi_\a = \psi_{ir,\a} = 
    \left(\begin{array}{cc} \psi_{i=1,r=1} & \psi_{i=1,r=2} \\
                         \psi_{i=2,r=1} & \psi_{i=2,r=2}\end{array}\right)_\a
= \left(\begin{array}{c} \psi_{i,r=1} \\
                         \psi_{i,r=2}\end{array}\right)_\a
\ee
cf. \eq{Dirac-spinor}, which
under the global $SU(2) \times SU(2)_R$  transforms as
\be
\Psi \to U \Psi U_R^T.
\ee
Now define
\be
\tilde\Psi = \sigma_2 \Psi \sigma_2,
\qquad \tilde \psi^{ir} = \varepsilon^{i j}\psi_{js}\varepsilon^{rs},
\ee
which using $U^{T} \sigma_2 U =\sigma_2$  transforms as
\be
\tilde\Psi \to \sigma_2 U \tilde\Psi U_R^T \sigma_2
= U^{-1 T} \sigma_2 \tilde\Psi \sigma_2 U_R^{-1}.
\ee
Note that $\sigma_2$ is the charge conjugation matrix
for $SU(2)$.
Then the mass  term can be written as
\be
S_m  = \int d^4 y\,  Tr\, \psi_{i,r,\a} 
\varepsilon^{\a\b}\varepsilon^{i j}\varepsilon^{rs} \psi_{j,s,\b} \,
 =  \int d^4 y\,  Tr\,\tilde\Psi_{\a}^T \varepsilon^{\a\b}\Psi_{\b}
\label{S-mass-1}
\ee
(+ h.c.).
This suggests to arrange the scalars similarly:
consider the antihermitian $2\cN \times 2\cN$ matrix
\be
\Phi = \phi_0\sigma_0 + \phi_a \sigma_a = \phi_\mu \sigma_\mu 
\label{Phi-extended}
\ee
including an additional component
$\phi_0$, which we set $\phi_0\equiv -\frac i2$ for now
but which will be allowed to be dynamical later.
The Yukawa coupling looks much nicer in this extended 
formalism:
\bea
S_Y  &=&  2i\int d^4 y\,  Tr\, \tilde\Psi_{\a}^T  \Phi\, \Psi_{\b}
\varepsilon^{\a\b}
 = \int d^4 y\,  Tr\, \tilde\Psi_{\a}^T  (\sigma_a\,[i\phi_a ,\Psi_{\b}] 
+ \{i\phi_0,\Psi_{\b}\}) \varepsilon^{\a\b} \nn\\
&=&  \int d^4 y\,  Tr\, \tilde\Psi_{\a}^T  \not \!\! D_{(2)}\,\Psi_{\b}
\varepsilon^{\a\b} ,
\label{S-yukawa-2}
\eea
where $ \not \!\! D_{(2)}$ is
precisely the Dirac operator on the fuzzy sphere,
with the correct constant shift due to $\phi_0 = -\frac i2$
(which will be understood naturally below). 
We can also impose the discrete symmetry
\be
\psi \to i\psi, \qquad \Phi \to -\Phi ,
\label{discrete-symm}
\ee
which excludes the mass term \eq{S-mass-1}, and 
requires the potential for  $\phi_\mu$  below to be 
even. This is very appealing, since the correct constant shift for the
Dirac operator on $S^2_N$ is then automatic, and no bare mass term is
allowed. It strongly suggests an underlying $SU(2\cN)$ structure.
However, 
the Yukawa coupling \eq{S-yukawa-2} 
 explicitly breaks $SU(2\cN)$ down to $SU(\cN) \times SU(2) \times
SU(2)_R$. Then there is 
another Yukawa coupling compatible with this unbroken symmetry,
\be 
\int d^4 y\,  Tr\, \tilde\Psi_{\a}^T  \phi_0\, \Psi_{\b}
\varepsilon^{\a\b} .
\ee 
This essentially amounts again to a mass term in our vacua, 
 spoiling to some
extent the reason for introducing the $SU(2\cN)$ structure. 
One could argue that  this term is absent at a fundamental
very high scale, and  will be induced
only  at lower scales due to renormalization,  
where this $SU(2\cN)$ symmetry is broken.

Further interesting possibilities appear once  we include another 
fermion $\kappa$ with the same properties as $\Psi$ as discussed in 
section \ref{sec:quantum-constraints}. This might then allow to
break $SU(2\cN)$ spontaneously. However, we leave such
explorations to future work.

\paragraph{Relation with twisted picture}

This extended formalism suggests to
consider the diagonal ``twisted'' $SU(2)_D\subset SU(2) \times SU(2)_R$ subgroup 
generated by $(U,U^{-1 T})$.
Decomposing $\Psi$ into vector and scalar
fermions
$\Psi = \Psi_V + \Psi_S$ under this $SU(2)_D$, where
\be
\Psi_{V,\a} = \psi_{a,\a} \sigma_a, 
\qquad \Psi_{S,\a} =  \psi_{0,\a} \sigma_0 ,
\ee
the Yukawa coupling
\eq{S-yukawa-2} can be rewritten as\footnote{Note that the transposition acts only on the 
$2\times 2$ block structure.}
\bea
S_Y  &=&  2i\int d^4 y\,  Tr\, \tilde\Psi_{\a}^T  \Phi\, \Psi_{\b}
\varepsilon^{\a\b} \nn\\
&=& i\int Tr\,  (-\Psi_{V,\a} \{\Phi, \Psi_{V,\b}\} 
- 2\Psi_{S,\a} \left[\Phi, \Psi_{V,\b}\right] 
+ \Psi_{S,\a} \{\Phi, \Psi_{S,\b}\})\varepsilon^{\a\b} 
\label{S-yukawa-4}
\eea
Now
\be
i\int Tr\,  \Psi_{V,\a} \{\Phi, \Psi_{V,\b}\} \varepsilon^{\a\b} 
= 2\int d^4 y\,  Tr\, \psi_{a,\a}  
\not \!\! \tilde D_{(2)}  \psi_{a,\b}\,\varepsilon^{\a\b} ,
\ee
where
\be
\(\not \!\! \tilde D_{(2)} \psi \)_{a,\a} 
= - \varepsilon_{abc} \, [\phi_b, \psi_{c,\a}] + \psi_{a,\a}
\ee
is the  ``vector-Dirac operator''.
Similarly, 
\be
 i\int d^4 y\, Tr\, \Psi_{S,\a} \left[\Phi, \Psi_{V,\b}\right] 
\varepsilon^{\a\b} 
= 2i\int d^4 y\, Tr\, \psi_{0,\a} \left[\phi_a,\psi_{a,\b}\right] 
\varepsilon^{\a\b} 
\label{scalar-vector-coupl}
\ee
and
\be
i\int d^4 y\, Tr\, \Psi_{S,\a} \{\Phi, \Psi_{S,\b}\}\varepsilon^{\a\b} 
= \int d^4 y\, Tr\, \psi_{0,\a}  \psi_{0,\b}\varepsilon^{\a\b} 
\ee
The kinetic term of the action is 
\bea
S_K &=& \int d^4 y\, Tr\, \Psi^\dagger 
i  \sigma^\mu(\partial_\mu + [A_\mu,.] ) \Psi  \nn\\
&=&  \int d^4 y\, Tr\, (\psi_{a,\a})^\dagger 
i {(\sigma^\mu)_{\dot\a}}^{\b} (\partial_\mu + [A_\mu,.]) \psi_{a,\b} 
 + (\psi_{0,\a})^\dagger 
i {(\sigma^\mu)_{\dot\a}}^{\b} (\partial_\mu + [A_\mu,.]) \psi_{0,\b}.
\nn
\eea
The 6-dimensional interpretation of this form is less obvious, 
because it looks like a vector on the internal sphere 
rather than a spinor.
As discussed in \cite{Andrews:2005cv}, it can be interpreted as a 
twisted compactification \cite{Maldacena:2000yy,Chamseddine:1997nm}, 
which is realized very naturally here.
The decomposition into KK modes and the 
low-energy effective action can be computed easily, by decomposing
these adjoint spinors into the eigenmodes of the ``vector-Dirac
operator''. 
We see that this is simply a different organization of our
doubled fermion picture.

A truly different model would be obtained if only
the twisted $SU(2)_D$ is a symmetry 
while the full $SU(2)\times SU(2)_R$ is broken. 
This would be consistent with additional
constraints such as $\Psi_0=0,\,[\Phi_a,\Psi_a] =0$, and
\be
\tilde\Psi = - \Psi^T.
\ee
However, this still does not give a
 chiral theory, since these constraints
are real.

\subsection{$SU(2\cN)$-extended scalar sector}

The above $SU(2\cN)$ formalism 
is very interesting for several reasons, and we discuss here how 
the scalar sector can be extended to be invariant under
this $SU(2\cN)$. 
Following \cite{Steinacker:2003sd,Steinacker:2007iq}, 
we consider the antihermitian $2\cN \times 2\cN$ matrix
\be
\Phi = \phi_0 \sigma_0 + \phi_a \sigma_a
\ee
as in \eq{Phi-extended},
including a scalar field $\phi_0 = -
\phi_0^\dagger$ in the adjoint of
$SU(\cN)$. For the moment we assume $\phi_0 = -\frac i2$, but 
$\phi_0$ will become dynamical below.
The main observation is  
\be
\Phi^2 = (\phi_a \phi_a + \phi_0\phi_0) \one + 
\frac 12 i \varepsilon_{abc} F_{bc} \sigma^a
\ee
with the generalized $S^2_N$ field strength
\be
F_{ab} = [\Phi_a,\Phi_b] - i\varepsilon_{abc} \{\phi_0,\phi_c\} .
\ee
We note in particular that any vacuum of the form 
\eq{solution-general}, we have
\be
\Phi^2 = (\phi_a \phi_a + \phi_0^2) \one_2 \propto \one_{2\cN}
\label{phi^2-one}
\ee
for a suitable $\phi_0 \approx -\frac i2$.
We can then furthermore impose the discrete symmetry 
\eq{discrete-symm}, 
which requires the potential to be even and 
excludes a mass term for the fermions. 

We now promote $\phi_0$ to a dynamical field, and
consider the extended $SU(2\cN)$ symmetry 
\be
\Phi \to U^{-1} \Phi U, \qquad U \in SU(2\cN) .
\label{su2N-symm}
\ee
The most general potential compatible with this symmetry is given by 
\be
V(\Phi) = Tr (M (\Phi \Phi + b^2\one)^2) \,\, + (\mbox{double-trace
  terms}) ,
\label{V-u2N}
\ee
where the double-trace terms\footnote{some of 
those would be eliminated by fixing the 
trace of $\Phi$, which however is not 
compatible with the discrete symmetry \eq{discrete-symm}.} have the form
$ c_1 Tr(\Phi^2) Tr(\Phi^2) + c_2 Tr(\Phi) Tr(\Phi^3) + c_3 (Tr(\Phi))^2
+ c_4 (Tr(\Phi))^4 + c_5 (Tr(\Phi))^2 Tr(\Phi^2) + c_6 (Tr(\Phi))^2$. 
The corresponding equation of motion has the form
$a\Phi^3 + b \Phi^2 + c\Phi + d =0$,
where the coefficients may involve trace terms. 
We assume furthermore that 
the term $M (\Phi \Phi + b^2\one)^2)$ dominates i.e. $M \to \infty$, 
while the double-trace
terms are naturally suppressed by $\frac 1{\cN}$. Then the vacuum
$\Phi$ decomposes into blocks which are  
small deformations of 
\be
\Phi^2 \propto \one,
\label{vac-2n-approx}
\ee
which as shown in \cite{Steinacker:2007iq} is 
naturally interpreted as fuzzy sphere with some
gauge group $U(n)$. 
All these solutions are degenerate with $V(\Phi)=0$ as long as 
$V(\Phi) = Tr (a (\Phi \Phi + b\one)^2)$. However, 
due to the presence of e.g. Yukawa terms which break the 
$SU(2\cN)$ symmetry, we expect after renormalization 
additional terms to be induced in the potential for $\phi_\a$ such as 
those appearing in \eq{pot}.
These as well as the double-trace terms of \eq{V-u2N} 
give different energy to different solutions 
with different block structure, completely analogous to the mechanism
discussed in section \ref{sec:vacua}. 
Again, it is very plausible that the same 
convexity argument generically leads to the same types of vacua
 with low-energy
$SU(n_1) \times SU(n_2) \times U(1)$ gauge symmetry.

\subsubsection{Constrained scalars}

In order to implement the chirality operator $\tilde\chi'$, 
it would be nice to impose a constraint of the form
\eq{vac-2n-approx}, which together with a suitable trace condition on
$Tr(\Phi)$ provides precisely the tangential 
gauge fields on the fuzzy sphere \cite{Steinacker:2007iq}.
A natural way to impose such a constraint is by adding 
the following renormalizable term to the action
\be
S = Tr M (\Phi^2 + c_\Lambda^2)^2 
\label{S-constraint}
\ee
and letting $M \to \infty$. This will indeed impose
the desired constraint $\Phi^2 = - c_\Lambda^2\,\one$
for $M \to \infty$, but we see that
a running of $c_\Lambda$ must be allowed.
Of course renormalization will induce other terms as well;
nevertheless one may hope that for $M\to \infty$,
the RG flow will only
change the eigenvalues of the projector, but not the property
that $\Phi$ has only 2 different eigenvalues.
A slightly different possibility is 
\be
S = Tr M (\Phi^2 + {c_\Lambda^2}'(Tr(\Phi))^2)^2 
\label{S-constraint-mod}
\ee
with $M \to \infty$, which amounts to 
the essentially equivalent constraint 
$\Phi^2 = - c_\Lambda(Tr(\Phi))^2 \one$.
This should be better behaved under
renormalization since only marginal operators occur.

Due to the presence of $SU(2\cN)$-breaking terms 
e.g. from the Yukawa terms, 
renormalization will induce additional terms, in 
particular\footnote{the precise values of the constants here are not essential} 
\be
S = -\frac {N}{g}\, Tr (\phi_0 + \frac i2)^2 .
\label{S-YM-alternative}
\ee
As shown in \cite{Steinacker:2007iq}, this
provides an alternative definition of Yang-Mills on 
the fuzzy sphere: $F = i\phi_0-\frac 12$ 
is the (scalar) 
field strength, while the constraint $\Phi^2 = -\frac 14\, N^2$ 
describes precisely 2 tangential gauge fields on $S^2_N$. 
This holds provided $Tr(\Phi) \sim  i N$, which we assume to 
follow from the double-trace terms in the action.
We therefore expect to find the same physics as discussed in the
previous sections. We refrain here from discussing the most general 
action compatible with the 
$SU(\cN) \times SU(2)\times SU(2)_R$ symmetry.

Nevertheless, implementing such a 
constraint in a 4D quantum field theory is far from trivial.
For example, taking $M \to \infty$ appears to be a  
strong coupling limit;
on the other hand, the term $(\Phi^2 + b^2)$ actually vanishes 
in the fuzzy sphere vacuum, and the 
desired modes consistent with this constraint are
not strongly coupled. 
Note also that this constraint 
essentially amounts to a certain type of
nonlinear sigma model in 4 dimensions, 
more precisely to projector-valued 
quantum fields (up to a shift). This provides renewed motivation
to study this type of field theory,
as well as an embedding in an extended SUSY model generalizing
\cite{Andrews:2005cv}.

Assuming such a constraint for $\Phi$, 
we can attempt to impose a chirality constraint  on the 
fermions.

\section{Chirality constraint: the quantum case}
\label{sec:quantum-constraints}

As we have seen explicitly in section \ref{sec:type2-zeromodes}, only in a chiral 
theory we can expect to get exactly massless fermions.
We recall the mechanism: $\Psi^{12}$ resp. 
$\Psi^{21}$ feel a $U(1)$ magnetic flux on $S^2$ with strength 
$k = N_1-N_2$ resp. $-k$ in the type II vacuum. 
This leads to $k$ would-be zero modes in $\Psi^{12}$ 
with positive chirality w.r.t. $S^2$, and 
$k$ would-be zero modes in $\Psi^{21}$ 
with negative chirality w.r.t. $S^2$. 
Thus in the chiral case 
only  $\Psi^{12}$ (or only  $\Psi^{21}$) is allowed,
and  there is no way for it to acquire a 4D mass without further 
symmetry breaking.
In a non-chiral case however, they can pair 
up and acquire a 4D mass.  

Imposing a chirality constraint on the 
quantum level turns out to be difficult, and we are not able to 
define a {\em renormalizable} 
model which is 6D chiral at the quantum level. 
The reason is that the 6D chirality operator is a dynamical
object which contains the scalar fields $\phi$. This must be so,
since at very high energies the model is again  4-dimensional,
which is the reason for maintaining renormalizability. 

Nevertheless, we discuss some strategies to impose
a 6D chirality constraint on the fermions.
The most promising approach is to use the modified 6D chirality
$\tilde\Gamma'$ \eq{chirality-constraint-prime} as discussed in section  \ref{subsec:chirality}.
For this to be well-defined at the quantum level, 
the constraint $\Phi^2 = c \one$ seems necessary. Then the definition
of $\tilde\Gamma'$ simplifies replacing $\tilde \Phi$ essentially by $\Phi$.
Taking into account possible renormalization
of $c_\Lambda$, the 6D chirality constraint $\tilde \tilde\Gamma'=\one$
\eq{chirality-constraint-prime}  becomes 
\be
(\Phi + c_\Lambda i\gamma_5)\Psi = 0
\label{PhiPsi-constraint}
\ee
or $(\Phi - c_\Lambda \, \gamma_5 Tr(\Phi))\Psi = 0$. 

There are 2 problems with this approach: first, $\Phi^2 \propto \one$
amounts to some kind of nonlinear sigma model in 
4 dimensions, which is not under control to our knowledge. Accordingly, it is not
clear if \eq{PhiPsi-constraint} can be imposed consistently on the
quantum level. The second problem is that the top modes of the
diagonal blocks apparently
become a large multiplet of massless fermions 
as discussed in section \ref{sec:type2-zeromodes}, 
which is clearly undesirable. 

\paragraph{Additional fermions.}

One might try to implement a similar constraint  by
including additional fermions $\kappa$, giving 
the ``wrong'' chirality of $\Psi$ a large mass.
Consider for example the action
\be
S_{\kappa} = Tr \tilde\kappa^T (\Phi \gamma_5 + c_\Lambda' Tr(\Phi)\,)\Psi
\, + S_{kin}(\kappa) + S_Y(\kappa) 
\label{Psi-chiralcoupl-1}
\ee
or\footnote{The main difference between \eq{Psi-chiralcoupl-1} and 
\eq{Psi-chiralcoupl-2} is that the latter  doesn't affect
the highest mode, which is in fact preferable as discussed in section 
\ref{subsec:chirality}}
\be
S_{\kappa} = Tr \tilde\kappa^T (\chi \gamma_5 + 1\,)\Psi
\, + S_{kin}(\kappa) + S_Y(\kappa) .
\label{Psi-chiralcoupl-2}
\ee
Here $\kappa$ are fermions 
in the adjoint of $SU(\cN)$ which
transforms under the global $SU(2) \times SU(2)_R$ as
\be
\kappa \to U_R^{-1 T} \kappa U^{-1}
\ee
Hence $\tilde \kappa^T \to U \tilde\kappa^T U_R^{T}$, and
all symmetries are preserved, including the discrete symmetry 
$\Phi \to -\Phi$.
This term couples all modes of $\Psi$ with the ``wrong'' 6D chirality to 
the ``opposite'' modes of $\kappa$ via terms of the form
$Tr\, c_n\,\rho_{\pm,(n),\a} \, \kappa^{\pm,(n),\a}$, giving them a 
large mass. 
While the kinetic term of $\kappa$ is essentially the same as for $\Psi$,
the Yukawa couplings can be chosen differently. 

In this situation,
those modes of $\Psi$ which satisfy the 6D chirality constraint 
do not (or only very weakly) couple to their counterparts in $\kappa$,
and essentially only their (would-be) zero modes survive. 
The surviving zero modes of 
$\Psi$ have a fixed 6D chirality, however those of $\kappa$ have the opposite 
chirality. This is quite similar to the non-chiral case, with the
additional feature that the Yukawa sector of $\Psi$ and $\kappa$
may be different. Again,
some fine-tuning is required 
to avoid these zero modes of $\Psi$ and $\kappa$ to
pair up and aquire a mass. This leads
again to 2 non-interacting or weakly-interacting
almost massless sets of fermions with opposite chirality and quantum
numbers. 

In any case, 
we end up essentially with a non-chiral model, which
requires (mild) fine-tuning in order to have 
approximately-massless
fermions due to the would-be zero modes. Those then come in 
``mirror pairs''  $\Psi^{12}$ resp. $\Psi^{21}$, 
i.e. an extra left-handed fermion for each right-handed one
with the opposite quantum numbers. 
Such a scenario is known as {\bf mirror fermions}, and
has been considered from the phenomenological 
point of view in \cite{Maalampi:1988va}. These models may become
chiral at low energies, since the mirror fermions
with the ``wrong'' chirality have different Yukawa sectors and are
assumed to be
heavier (of the order of the 
electroweak scale \cite{Maalampi:1988va}) and hence
hidden at low energies.

There are other constraints which could be imposed on 
the fermions, one of which is described below. 
However they do not appear
to give complex chiral low-energy fermions.

\paragraph{Alternative constraints}

One may try to impose a 
``twisted'' version of chirality, in terms of
\be
\tilde R\Psi  := \frac{2i}N\,\Psi\, \Phi,
\label{chiral-proj-twist}
\ee
assuming again $\Phi^2 \sim \one$.
Then imposing a twisted chirality $\tilde\chi' \tilde R =1$ amounts to
\be
(\tilde\chi'\tilde R -1)\Psi  = 0 \quad\Longleftrightarrow \quad \Phi \Psi = \Psi \Phi
\ee
while 
\be
(\tilde\chi' \tilde R +1)\Psi = 0 \quad\Longleftrightarrow \quad \Phi \Psi = -\Psi \Phi .
\ee
This relates $SU(2)_V$ to $SU(2)_R$,
and preserves only the diagonal $SU(2)_D$ of the
twisted picture in section \ref{fermions-extended}.
Such a  constraint could be imposed more easily 
on the quantum level, 
by adding 
\be
Tr M\, \Psi_\a \{\Phi, \Psi_\b\}\,\varepsilon^{\a\b}
\label{alternate-constr}
\ee
to the action with large $M$. 
However, our (in-exhaustive) 
analysis of this and similar constraints 
did not reveal a chiral low-energy theory.
Note that \eq{alternate-constr} is 
preserved under renormalization; 
imposing e.g. \eq{PhiPsi-constraint} or $\tilde R\Psi=\Psi$  is much more difficult.

\section{Fluxons and prospects for low-energy SSB}
\label{sec:fluxons-ssb}

We suggest here briefly a possible mechanism for further 
(low-energy, electroweak) symmetry breaking. This is speculative at
this point, nevertheless it is compelling and natural enough to justify further 
investigation.

Consider a type II vacuum with an additional fluxon present.
Then the scalars for the vacuum can be 
\be 
\phi_a = \left(\begin{array}{ccc} \a_1\,
X_a^{(N_1)}\otimes\one_{n_1}& 0 &0\\ 0 &
\a_2\,X_a^{(N_2)}\otimes\one_{n_2} & D_a \\ 0 & -D_a^\dagger
& c_a
             \end{array}\right),
\label{vacuum-mod5}
\ee
where $c_a \in i\R$ denotes the position of the fluxon on $S^2$, 
and  we assume furthermore a nontrivial off-diagonal column
$D_a$. This  is the key to further SSB.
To establish (or exclude) 
such a vacuum would require more detailed analysis of the 
scalar potential $V(\phi_i)$ which has not been attempted.
The crucial point is that any $D_a \neq 0$ can be transformed in the
form $D_a^\dagger = (0, ...,0, d_a^*)$, which
implies that 
$SU(n_2)$ is broken spontaneously to $SU(n_2-1)$. 
Assuming that  $n_1=3, n_2=2$ this amounts to further 
breaking 
$SU(3) \times SU(2) \times U(1)\times U(1) \to SU(3) \times U(1) \times
U(1)$, where
$D_a$ plays the role of the ``electroweak'' Higgs.
For the fermions
we expect zero modes in the off-diagonal blocks,
\be
\psi_{eff} = \left(\begin{array}{ccc} 0 & \psi_{32} & \psi_{31}\\
 \psi_{23} & 0 & \psi_{21} \\
\psi_{13} & \psi_{12} & 0 \end{array}\right)
\ee
while the fluxon sectors are localized on $S^2$ and therefore
essentially 4-dimensional. 
In particular,  $\left(\begin{array}{c}
    \psi_{31}\\ \psi_{21}\end{array}\right)$ 
corresponds to a fundamental $(5)\to (3) \oplus (2)$ of 
$SU(5) \to SU(3) \times SU(2)$, providing 
right-handed quarks and left-handed leptons.  $\psi_{32}$
is in the bifundamental of $SU(3) \times SU(2)$ 
corresponding to left-handed quarks. The right-handed leptons might 
arise from diagonal fluxon block, but this is purely speculative
at this point.
The family number should arise from the index $k$, which however
appears to prefer $k=1$. In any case at present 
this is just a toy model, with the aim to establish 
the basic mechanisms.

\subsubsection*{Relation with CSDR scheme}

It is interesting to look at the  results of this paper 
from the point of view of coset space dimensional reduction (CSDR).
In \cite{Aschieri:2003vy}, similar 
effective 4-dimensional models
are constructed starting from gauge theory on 
$M^4 \times S^2_N$, by imposing  
CSDR constraints following the general ideas of 
\cite{Forgacs:1979zs,Kapetanakis:1992hf}. 
These constraints
boil down to choosing embeddings  of $SU(2)
\subset SU(\cN)$, which can be identified with 
 the possible block configurations \eq{solution-general}.  
The solutions of the constraints can be
formally identified with the lowest modes of the KK-towers 
of the fields.  
On the other hand, 
the present approach takes into account the most 
general renormalized  potential, leading to a 
vacuum selection mechanism and nontrivial fluxes.

In the context of ordinary CSDR, the question of chiral fermions 
has been studied in \cite{Chapline:1982wy,Barnes:1986ea}.
This appears to be consistent with our conclusion that the 
model in the present setting is non-chiral. 
Nevertheless, it is not entirely
clear to which extent these commutative results are applicable to the
fuzzy case.

\section{Discussion}

We have presented a simple,
renormalizable 4-dimensional $SU(\cN)$ 
gauge theory with suitable 
scalar and fermionic matter content, which  spontaneously
develops an extra-dimensional fuzzy sphere. 
The underlying mechanism is simply SSB and the Higgs effect.
 The model behaves as a 6-dimensional
 Yang-Mills theory on $M^4 \times S^2_N$, for energies below
a cutoff $\Lambda_{6D} = \frac{N^2}R$. 
The expected KK modes for the fermions are found, 
extending the bosonic analysis 
given in \cite{Aschieri:2006uw}. 
This model is remarkable not only for this
striking behavior, but also for 
a natural mechanism for obtaining an
unbroken gauge group $SU(n_1) \times SU(n_2) \times U(1)$
as well as zero modes due to
a magnetic flux on $S^2_N$. It represents a particularly
simple yet rich realization of the idea of deconstructing dimensions
\cite{Arkani-Hamed:2001ca}, taking advantage of 
results from noncommutative field theory.
This allows to consider ideas of compactification and 
dimensional reduction within a renormalizable framework.
Our framework provides in fact a slight generalization
of the conventional geometric compactification, which manifests itself e.g. in
the different effective radii seen by fermions and gauge fields.
Moreover, using the results of \cite{Steinacker:2007dq}
this mechanism can be understood as an effect of gravity in 
extra dimensions.

However, it turns out that the model is 
 non-chiral a priori, and imposing a
chirality constraint appears to be very difficult 
on the quantum level.
This means that each would-be zero mode from $\Psi^{12}$ 
has a mirror partner from $\Psi^{21}$, with opposite
chirality and gauge quantum numbers. 
Thus we arrive essentially at a picture of mirror fermions 
discussed e.g. in \cite{Maalampi:1988va} from a 
phenomenological point of  view.
While this may still be interesting 
physically since the ``mirror fermions'' 
may have larger mass as the ones we see at low energies,
it would be desirable to find a chiral version 
with similar features. There are indeed many possible 
directions for generalizations, exploring other types
of fuzzy internal spaces. We hope to report on 
such an extension to $\C P^n_N$ soon.

Another interesting generalization would be supersymmetry. 
This is natural since both bosons and fermions are in the 
adjoint of the gauge group; furthermore,
the observation of section \ref{sec:SU2N} 
that both the fermions and  scalars 
are naturally arranged as adjoint of 
$SU(2\cN)$ points to 
supersymmetry at some higher scale.
Indeed, a very similar SUSY model has already been 
discussed in  \cite{Andrews:2005cv} 
which also develops an extra-dimensional sphere.
That model is related to the Maldacena-
Nu\~nez twisted compactification.
However, our mechanism for vacuum selection and 
obtaining a type II vacuum
unbroken gauge group $SU(n_1) \times SU(n_2) \times U(1)$
no longer applies in that model, since SUSY is unbroken. 
 This suggests to search for a
SUSY version of our model, where the nontrivial 
type II vacua are
accompanied by spontaneous SUSY breaking.

Finally, a natural generalization of this idea is to spontaneously
generate not only the extra dimensions but also the ``visible'' ones.
This leads to the matrix-model approach to noncommutative gauge
theory, which at least in the Euclidean case 
has been elaborated in several
models such as \cite{Grosse:2004wm}, or e.g.
 \cite{Azuma:2004qe} for 
matrix models related to string theory.
Taking into account results in \cite{Abel:2005rh},
a combination of such models with fuzzy extra dimensions might be
particularly interesting.

\paragraph{Acknowledgments}

We are grateful for discussions with H.~Aoki, P. Aschieri, S. Bal, 
A.P. Balachandran, W. Grimus,
H. Grosse, C.-S. Chu, X. Martin, S. Nicolis, J. Nishimura, 
 U. and S. Watamura.
This work is supported by the EPEAEK programme "PythagorasII" and 
co-founded by the European Union (75\%) 
and the Hellenic State (25\%).  
The work of H.S. is supported by the FWF under project
P18657.

\section{Appendix}

\subsection*{Appendix 1: The fuzzy sphere}
\label{sec:fuzzysphere}

The fuzzy sphere \cite{Madore:1991bw} is a matrix approximation of the
usual sphere $S^2$. The algebra of functions on $S^2$ (which is
spanned by the spherical harmonics) is truncated at a given frequency
and thus becomes finite dimensional.  The algebra then becomes that of
$N\times N$ matrices. More precisely, the algebra of functions on the
ordinary sphere can be generated by the coordinates of $\R^3$ modulo
the relation $ \sum_{ {a}=1}^{3} {x}_{ {a}}{x}_{ {a}} =r^{2}$. The
fuzzy sphere $S^2_{N}$ is the non-commutative manifold whose
coordinate functions 
\be {x}_{ {a}} = r\,\frac {i}{\sqrt{C_2(N)}}\, {X}_{
{a}}, \qquad {x}_{ {a}}^\dagger = {x}_{ {a}}
\label{fuzzycoords}
\ee
are $N \times N$ hermitian matrices proportional to the generators of
the $N$-dimensional representation of $SU(2)$. They satisfy the
condition $\sum_{{a}=1}^{3} x_{{a}} x_{{a}} = r^2$ and the commutation
relations
\begin{equation}
[ X_{{a}}, X_{{b}} ] = \varepsilon_{abc}\, X_{{c}}~.
\end{equation}
For $N\rightarrow \infty$, one recovers the usual commutative sphere.
The best way to see this is to decompose the space of functions on
$S^2_{N}$ into irreps under the $SU(2)$ rotations, \bea S^2_{N} \cong
(N) \otimes (N) &=& (1) \oplus (3) \oplus ... \oplus (2N-1) \nn\\ &=&
\{Y^{0,0}\} \,\oplus \, ... \, \oplus\, \{Y^{(N-1),m}\}.
\label{fuzzyharmonics}
\eea This provides at the same time the definition of the fuzzy
spherical harmonics $Y^{lm}$, which we normalize as \be Tr_N
\left((Y^{lm})^\dagger Y^{l'm'}\right) = \delta^{l l'} \delta^{m m'}.
\ee Furthermore, there is a natural $SU(2)$ covariant differential
calculus on the fuzzy sphere. This calculus is three-dimensional, and
the derivations of a function $ f$ along $X_{{a}}$ are given by
$e_{{a}}({f})=[X_{{a}}, {f}]\,.\label{derivations}$ These are
essentially the angular momentum operators
\begin{equation}\label{LDA}
 J_a f = i{e}_{{a}} f = [i{X}_{{a}},f ],
\end{equation}
which satisfy the $SU(2)$ Lie algebra relation
 \begin{equation}
[J_a, J_b ] = i\varepsilon_{abc} J_c.
 \end{equation}
In the $N \rightarrow \infty$ limit the derivations $e_{{a}}$ become
$e_{{a}} = \varepsilon_{abc} x_{{b}}\partial_{{c}}$, and only in this
commutative limit the tangent space becomes two-dimensional. 
For further developments see
e.g. \cite{Balachandran:2005ew,Grosse:1995ar,Chu:2001xi}
and references therein.

\subsection*{Appendix 2: The spectrum of $\not \!\! D_{(2)}$}
\label{sec:append-spec}

Let us work out the spectrum of $\not \!\! D_{(2)}$ in detail.
For $\phi_a$ given by \eq{vacuum-mod1}, we have 
\bea
\not \!\! D_{(2)} \Psi
&=&i\sigma_a (\phi_a \Psi - \Psi \phi_a)  + \Psi \nn\\
&=& (\a \sigma_a J_a + 1)\Psi
 = \a (C_2- \frac 34 - J^2) \Psi +  \Psi ,
\eea
where $C_2 := (\frac 12 \sigma_a + J_a)^2$ is the quadratic Casimir.
Ignoring the extra $\msu(n)$ degrees of freedom, this can be evaluated
on the decomposition \eq{spinor-decomp} using some $SU(2)$ algebra. 
The eigenvalues of $\not \!\! D_{(2)}$ on the modes $\Psi_{\pm,(n)}$ are given
by 
\be \not \!\! D_{(2)}\Psi_{\pm,(n)} = \left(\a (C_2- \frac 34 - J^2)
+ 1\right) \Psi_{\pm,(n)} ,
\ee 
where $C_2 = \frac 14(n^2-1)$, $J^2 =
\frac 14((n\mp 1)^2-1)$, and thus 
\bea \not \!\! D_{(2)}\Psi_{\pm,(n)}
&=& \left(\frac{\a}4 (n^2 -(n\mp 1)^2-3) + 1\right) \Psi_{\pm,(n)}\nn\\
&=& \left(\pm \frac{\a}2 n + (1-\a)\right) \Psi_{\pm,(n)} \, 
= E_{\delta = \pm,(n)}\,\Psi_{\pm,(n)} \,  
\label{D-spectrum-comp}
\eea 
with
\be
E_{\delta = \pm,(n)}\, \approx\, \frac{\a}2\, \left\{\begin{array}{rl} 
n, & \delta = 1,\qquad n=2,4,..., 2N \\
-n, & \delta = -1,\quad\, n=2,4,..., 2N-2
\end{array}\right.
\ee
assuming $\a \approx 1$; this is exact for $\a =1$. This can easily be
generalized to the type II vacuum, which is not needed however.

Th eigenvalue of $\chi$ can be worked out similarly using
\be
 \chi(\Psi^\pm) = \frac{1}{N} \sigma_a (i\phi_a^L + i\phi_a^R)
 = \frac{1}{N}\left(\a_1 J_{1L}^2 - \a_2 J_{1R}^2  
 - \a_1^2 X_L^2 + \a_2^2 X_R^2  \right) \Psi_{\pm,n},
\label{chi-explicit}
\ee
where $C_2 = \frac 14(n^2-1)$, 
$J_{1L,a} = \frac 12 \sigma_a + X_a^L$, and 
$J_{1R,a} = \frac 12 \sigma_a - X_a^R$. This is written for the case
of the type II vacuum.
In particular, for the highest mode of the diagonal blocks we have
$J_{1L}^2 = \frac 14((N+1)^2-1) = J_{1R}^2$, hence
$\chi(\Psi^{12}_{+,(2N)}) = 0$ exactly. 

A quick way to determine the chirality for the lowest modes is
indicated in the main text.

\paragraph{2D chirality of the conjugate spinors}

Recall from \eq{Dirac-spinor} that 
$\obar{\rho}_i^{\dot\a} = \varepsilon_{ij} \, \varepsilon^{\dot \a\dot\b}\, 
(\rho_{\b,i})^\dagger$, hence 
$\obar{\rho}^{\dot\a} = (i \sigma_2) \, \varepsilon^{\dot \a\dot\b}\, 
(\rho_{\b})^{\dagger T_2}$ where $T_2$ denotes transposition of the 
matrix indices $i,j$. Then 
consider
\bea
\obar{\chi\rho}^{\dot\a} &=&  \varepsilon^{\dot \a\dot\b} (i \sigma_2)
(\chi\rho_\b)^{\dagger T_2} 
= -\frac iN\, \varepsilon^{\dot \a\dot\b} (i \sigma_2)
(\sigma_a\{\Phi_a,\rho_\b\})^{\dagger T_2} \nn\\
&=& \frac iN\, \varepsilon^{\dot \a\dot\b} (i \sigma_2)
\sigma_a^{\dagger T_2}\{\Phi_a,\rho_\b^{\dagger T_2}\} \nn\\
&=& \frac iN\, \varepsilon^{\dot \a\dot\b} (i \sigma_2)
(-\sigma_2\sigma_a\sigma_2)\,(\{\Phi_a,\rho_\b^{\dagger T_2}\}) \nn\\
&=& -\frac iN\, 
\sigma_a \varepsilon^{\dot \a\dot\b}
(i\sigma_2)\,(\{\Phi_a,\rho_\b^{\dagger T_2}\}) \nn\\
&=& - \chi(\obar\rho^{\dot\a}) ,
\eea
where we used $\sigma_a^T = - \sigma_2 \sigma_a \sigma_2$ and 
antihermiticity of $\phi_a$.

For the zero modes, this can be understood by noting that 
$\rho \in (2)\otimes (N+m) \otimes (N)$ while 
$\obar\rho \in (2)\otimes (N) \otimes (N+m)$, since the transposition
$T_2$ acts only on the spinor indices.

\bibliographystyle{diss}

\bibliography{mainbib}

\begin{thebibliography}{99}

\bibitem{Arkani-Hamed:2001ca}
  N.~Arkani-Hamed, A.~G.~Cohen and H.~Georgi,
  ``(De)constructing dimensions,''
  Phys.\ Rev.\ Lett.\  {\bf 86} (2001) 4757
  [hep-th/0104005].

\bibitem{Aschieri:2006uw}
  P.~Aschieri, T.~Grammatikopoulos, H.~Steinacker and G.~Zoupanos,
  ``Dynamical generation of fuzzy extra dimensions, dimensional reduction and
  symmetry breaking,''
  JHEP {\bf 0609} (2006) 026
  [hep-th/0606021];
  P.~Aschieri, H.~Steinacker, J.~Madore, P.~Manousselis and G.~Zoupanos,
  ``Fuzzy Extra Dimensions: Dimensional Reduction, Dynamical Generation and
  Renormalizability,''
  [arXiv:0704.2880].


\bibitem{Aschieri:2003vy}
  P.~Aschieri, J.~Madore, P.~Manousselis and G.~Zoupanos,
  ``Dimensional reduction over fuzzy coset spaces,''
  JHEP {\bf 0404}, 034 (2004)
  [hep-th/0310072];
  P.~Aschieri, J.~Madore, P.~Manousselis and G.~Zoupanos,
  ``Renormalizable theories from fuzzy higher dimensions,''
  [hep-th/0503039];
P.~Aschieri, J.~Madore, P.~Manousselis and G.~Zoupanos,
  ``Unified theories from fuzzy extra dimensions,''
  Fortsch.\ Phys.\  {\bf 52}, 718 (2004)
  [hep-th/0401200].


\bibitem{Steinacker:2007dq}
  H.~Steinacker,
  ``Emergent Gravity from Noncommutative Gauge Theory,''
  arXiv:0708.2426 [hep-th].


\bibitem{Maalampi:1988va} J. Maalampi, M. Roos, ``The physics of Mirror fermions''.
{\em Phys.Rept.} {\bf 186} :53,1990.


\bibitem{Dolan:2002ck}
  B.~P.~Dolan and C.~Nash,
  ``The standard model fermion spectrum from 
 complex projective spaces,''
  JHEP {\bf 0210} (2002) 041
  [hep-th/0207078];
  B.~P.~Dolan,
  ``Non-commutative complex projective spaces and the standard model,''
  Mod.\ Phys.\ Lett.\  A {\bf 18} (2003) 2319
  [hep-th/0307124].

\bibitem{Aoki:2004sd}
H.~Aoki, S.~Iso, T.~Maeda and K.~Nagao,
  ``Dynamical generation of a nontrivial index on the fuzzy 2-sphere,''
  Phys.\ Rev.\  D {\bf 71} (2005) 045017
  [Erratum-ibid.\  D {\bf 71} (2005) 069905]
  [hep-th/0412052].

\bibitem{Aoki:2006wv}
  H.~Aoki, S.~Iso and T.~Maeda,
  ``On the Ginsparg-Wilson Dirac operator in the monopole backgrounds on the fuzzy 2-sphere,''
  Phys.\ Rev.\  D {\bf 75} (2007) 085021
  [hep-th/0610125].


\bibitem{Andrews:2005cv}
  R.~P.~Andrews and N.~Dorey,
  ``Spherical deconstruction,''
  Phys.\ Lett.\  B {\bf 631}, 74 (2005)
  [hep-th/0505107];
  R.~P.~Andrews and N.~Dorey,
  ``Deconstruction of the Maldacena-Nunez compactification,''
  Nucl.\ Phys.\  B {\bf 751}, 304 (2006)
  [hep-th/0601098].


\bibitem{Steinacker:2003sd}
  H.~Steinacker,
  ``Quantized gauge theory on the fuzzy sphere as random matrix model,''
  Nucl.\ Phys.\  B {\bf 679}, 66 (2004)
  [hep-th/0307075];
  H.~Steinacker,
  ``Gauge theory on the fuzzy sphere and random matrices,''
  [hep-th/0409235].

\bibitem{Carow-Watamura:1998jn}
  U.~Carow-Watamura and S.~Watamura,
  ``Noncommutative geometry and gauge theory on fuzzy sphere,''
  Commun.\ Math.\ Phys.\  {\bf 212}, 395 (2000)
  [hep-th/9801195].


\bibitem{Presnajder:2003ak}
  P.~Presnajder,
  ``Gauge fields on the fuzzy sphere,''
  Mod.\ Phys.\ Lett.\  A {\bf 18}, 2431 (2003).

\bibitem{Steinacker:2007iq}
  H.~Steinacker and R.~J.~Szabo,
  ``Localization for Yang-Mills theory on the fuzzy sphere,''
  [hep-th/0701041].

\bibitem{Grosse:1994ed}
  H.~Grosse and P.~Presnajder,
  ``The Dirac operator on the fuzzy sphere,''
  Lett.\ Math.\ Phys.\  {\bf 33}, 171 (1995).

\bibitem{Carow-Watamura:1996wg}
  U.~Carow-Watamura and S.~Watamura,
  ``Chirality and Dirac operator on noncommutative sphere,''
  Commun.\ Math.\ Phys.\  {\bf 183} (1997) 365
  [hep-th/9605003].

\bibitem{Balachandran:2000du}
  A.~P.~Balachandran, T.~R.~Govindarajan and B.~Ydri,
  ``The fermion doubling problem and noncommutative geometry. II,''
  [hep-th/0006216];

\bibitem{Balachandran:2003ay}
  A.~P.~Balachandran and G.~Immirzi,
  ``The fuzzy Ginsparg-Wilson algebra: 
A solution of the fermion doubling problem,''
  Phys.\ Rev.\  D {\bf 68}, 065023 (2003)
  [hep-th/0301242].


\bibitem{Maldacena:2000yy}
  J.~M.~Maldacena and C.~Nu\~nez,
  ``Towards the large N limit of pure N = 1 super Yang Mills,''
  Phys.\ Rev.\ Lett.\  {\bf 86} (2001) 588
  [hep-th/0008001].

\bibitem{Chamseddine:1997nm}
A.~H.~Chamseddine and M.~S.~Volkov,
  ``Non-Abelian BPS monopoles in N = 4 gauged supergravity,''
  Phys.\ Rev.\ Lett.\  {\bf 79} (1997) 3343
  [arXiv:hep-th/9707176];
A.~H.~Chamseddine and M.~S.~Volkov,
  Phys.\ Rev.\  D {\bf 57} (1998) 6242
  [arXiv:hep-th/9711181].


\bibitem{Grosse:2004wm}
  H.~Grosse and H.~Steinacker,
  ``Finite gauge theory on fuzzy $\C P^2$,''
  Nucl.\ Phys.\  B {\bf 707}, 145 (2005)
  [hep-th/0407089];
 W.~Behr, F.~Meyer and H.~Steinacker,
  ``Gauge theory on fuzzy $S^2 \times S^2$ and 
regularization on noncommutative $\R^4$,''
  JHEP {\bf 0507}, 040 (2005)
  [hep-th/0503041].


\bibitem{Azuma:2004qe}
  T.~Azuma, S.~Bal, K.~Nagao and J.~Nishimura,
  ``Dynamical aspects of the fuzzy $\C P^2$ in the large N reduced model with  a
  cubic term,''
  JHEP {\bf 0605} (2006) 061
  [hep-th/0405277];
  T.~Azuma, S.~Bal, K.~Nagao and J.~Nishimura,
  ``Nonperturbative studies of fuzzy spheres in a matrix model with the
  Chern-Simons term,''
  JHEP {\bf 0405} (2004) 005
  [hep-th/0401038];
  T.~Azuma, S.~Bal and J.~Nishimura,
  ``Dynamical generation of gauge groups in the massive
  Yang-Mills-Chern-Simons matrix model,''
  Phys.\ Rev.\  D {\bf 72} (2005) 066005
  [hep-th/0504217].


\bibitem{Abel:2005rh}
  S.~A.~Abel, J.~Jaeckel, V.~V.~Khoze and A.~Ringwald,
  ``Noncommutativity, extra dimensions, and power law running in the
  infrared,''
  JHEP {\bf 0601} (2006) 105
  [hep-ph/0511197].


\bibitem{Madore:1991bw}
  J.~Madore,
  ``The fuzzy sphere,''
  Class.\ Quant.\ Grav.\  {\bf 9} (1992) 69.


\bibitem{Grosse:1995ar}
  H.~Grosse, C.~Klimcik and P.~Presnajder,
  ``Towards Finite Quantum Field Theory In Noncommutative Geometry,''
  Int.\ J.\ Theor.\ Phys.\  {\bf 35}, 231 (1996)
  [hep-th/9505175].

\bibitem{Chu:2001xi}
  C.~S.~Chu, J.~Madore and H.~Steinacker,
  ``Scaling limits of the fuzzy sphere at one loop,''
  JHEP {\bf 0108}, 038 (2001)
  [hep-th/0106205].


\bibitem{Balachandran:2005ew}
  A.~P.~Balachandran, S.~Kurkcuoglu and S.~Vaidya,
  ``Lectures on fuzzy and fuzzy SUSY physics,''
  [hep-th/0511114].


\bibitem{Forgacs:1979zs}
  P.~Forgacs and N.~S.~Manton,
  ``Space-Time Symmetries In Gauge Theories,''
  Commun.\ Math.\ Phys.\  {\bf 72} (1980) 15.

\bibitem{Kapetanakis:1992hf}
  D.~Kapetanakis and G.~Zoupanos,
  ``Coset Space Dimensional Reduction Of Gauge Theories,''
  Phys.\ Rept.\  {\bf 219}, 1 (1992).

\bibitem{Chapline:1982wy}
  G.~Chapline and R.~Slansky,
  ``Dimensional Reduction And Flavor Chirality,''
  Nucl.\ Phys.\  B {\bf 209}, 461 (1982).

\bibitem{Barnes:1986ea}
  K.~J.~Barnes, P.~Forgacs, M.~Surridge and G.~Zoupanos,
  ``On Fermion Masses in a Dimensional Reduction Scheme,''
  Z.\ Phys.\  C {\bf 33}, 427 (1987).




\end{thebibliography}

\end{document}